\documentclass[twocolumn]{aa}
\usepackage{graphicx} 
\usepackage{color}
\usepackage{amsmath,graphicx,amssymb}
\usepackage[varg]{txfonts}
\usepackage{natbib}
\usepackage[switch]{lineno}
\usepackage{xcolor}
\usepackage{media9}
\usepackage{mathrsfs}  
\usepackage[T1]{fontenc}
\usepackage{comment}
\usepackage{hyperref}
\usepackage{placeins}
\usepackage{threeparttable}
\usepackage{multirow}
\definecolor{DarkBlue}{rgb}{0.0,0.1,0.4}
\definecolor{Red}{rgb}{0.6,0.2,0.1}
\definecolor{DarkRed}{rgb}{0.5,0.0,0.5}
\definecolor{Blue}{rgb}{0.0,0.0,1.0}
\definecolor{Zelinkava}{rgb}{0.2,0.5,0.2}
\definecolor{Pink}{rgb}{0.0,0.7,0.7}
\definecolor{White}{rgb}{1.0,1.0,1.0}
\bibpunct{(}{)}{;}{a}{}{,}

\hypersetup{
    colorlinks=true,
    linkcolor=Blue,
    citecolor=Blue,
    filecolor=Blue,     
    urlcolor=Blue,
}

\defcitealias{PK19}{PK19}
\defcitealias{PK20}{PK20}

\title{Supernova interactions with aspherical circumstellar
material I: calculations of light curves, AB magnitudes, spectra, and polarisation}
\author{P.~Kurf\"urst\inst{1} \and G.~Bless\inst{1} \and 
J. Fi\v s\' ak\inst{2}\and F.~Holoubek\inst{1} \and J.~Krti\v{c}ka\inst{1} \and B.~Kub\'{a}tov\'{a}\inst{2} \and J.~Kub\'{a}t\inst{2} \and M.~Zaja\v{c}ek\inst{1}}

\institute{Department of Theoretical Physics and Astrophysics,
           Masaryk University, Kotl\' a\v rsk\' a 2, 611 37 Brno, Czech Republic, \email{petrk@physics.muni.cz}\and Astronomical Institute of the Czech Academy of Sciences, Fričova 298, 251 65 Ondřejov, Czech Republic}
          
\date{Received}

\begin{document}

\abstract{We present an upgraded detailed numerical calculations of supernova (SN) interactions with significantly aspherical circumstellar matter (CSM), primarily formed as a disc or bipolar lobes. The circumstellar disc can arise as a result of, for example, mass transfer in a binary, while bipolar lobes can be the result of a violent pre-explosive ejection of matter, similar to the iconic cases of luminous blue variable stars (LBVs).
We numerically simulate the radiation-hydrodynamic (RHD) behaviour of interaction processes using a 2D cylindrical version of the RHD code CASTRO. We then calculate light curves, spectral patterns, and polarisation profiles, all up to a relatively long time of two years after an SN shock breakout and from different directions, using the multidimensional Monte Carlo radiation transfer (MC-RT) codes SEDONA and SIROCCO.
We calculated a total of five models for the two aforementioned configurations of the surrounding CSM, for stratified density levels, comparing the simulated hydrodynamic behaviour and differences in their observable properties. RHD models exhibit similar behaviour to previous adiabatic models, but with a significantly slower expansion velocity. The calculated light curves show a relatively smooth evolution in SN-disc interaction, and declines and brightening in SN-lobes interaction. Comparing models with real events with a presumed similar physical process provides guidance for selecting a more accurate CSM configuration when simulating real situations.}


\titlerunning{Supernova explosion long term interactions with aspherical circumstellar
material}
\authorrunning{P.~Kurf\"urst et al.}
\maketitle

\section{Introduction}
\label{intro}
Interactions of supernovae (SNe) with sufficiently dense circumstellar matter (CSM) can significantly alter the 
evolution of light curves (LCs) and other observable characteristics, such as spectra and polarisation properties; see e.g. \citet{1984ApJ...285L..63C}, \citet{1994ApJ...420..268C}, \citet{2006ApJ...651..381C}, and \citet{2003LNP...598..171C}, as well as \citet{2020RSOS....700467F} for reviews. 
In most notable cases, this process gives rise to 
the super-luminous SNe \citep[SLSNe, e.g.,][]{2007ApJ...668L..99Q,GalYam2012}. These can be $1\,\text{\textendash}\,2$ magnitudes brighter at maximum than the brightest SNe powered only by the progenitor's thermal energy and the 
radioactive decay of newly synthesised elements; although this may not be the only scenario for producing such extraordinary luminosity \citep[e.g.,][]{2010ApJ...719L.204W,2010ApJ...717..245K,2013Natur.502..346N,2015MNRAS.454.3311M}.
Another observed feature that reveals SNe interacting with their surroundings is the occurrence of narrow spectral lines, at least in certain phases; these SNe are classified as type IIn \citep{1990MNRAS.244..269S}.

We perform numerically consistent 2D simulations of axisymmetric radiation hydrodynamics (RHD) of spherical SN interacting with CSM with two aspherical morphologies 
\citep[cf.][hereafter \citetalias{PK20}]{2018ApJ...856...29M,PK20}: a circumstellar disc and a system of bipolar lobes, scaled in individual models with different initial densities. 
Based on numerically calculated observables, our goal is to identify  qualitative differences between the various CSM morphologies, in combination with their density.
Once such a correlation is convincingly found, it can contribute to a more precise determination of the mass loss rate of massive stars before the SN \citep[e.g.,][]{2014ARA&A..52..487S}. It may also determine the spatial orientation of ejected material during violent eruptions in the pre-SN period or the rate of mass transfer between components in binary systems \citep[cf., e.g.,][]{2020A&A...640A..33Z}. 

The analysis of the SN ejecta interaction with the surrounding medium is also relevant in the context of denser galactic environments, such as molecular clouds and galactic nuclei. Previous studies of the SN interaction with molecular clouds \citep{1999ApJ...511..798C,2015SSRv..188..187S} pointed out their X-ray, radio, and $\gamma$-ray signatures and the relevance in terms of cosmic ray contribution via diffuse shock acceleration. One of the key signatures of the interaction of the propagating SN shock with the molecular cloud is the compact OH (1720\,MHz) maser emission \citep{1996AJ....111.1651F}. The SN interaction with AGN accretion discs is expected to lead to luminous X-ray/UV/optical flares that could have been mistaken for some tidal disruption events \citep{2021MNRAS.507..156G,2023ApJ...950..161L}. In the context of star-jet collisions \citep{2020ApJ...903..140Z,2025MNRAS.540.1586K}, the SN-jet interaction \citep{2019A&A...622A.175V,2023A&A...677L..14B,2025A&A...704A.172L} can be the source of high energy $\gamma$-ray emission and can lead to the production of ultra-high-energy cosmic rays. 

Since this work primarily seeks to explore the relationships between the observables during SN expansion and the various morphologies and densities of pre-SN CSMs, we do not create specific models of individual SN events here. However, we compare the theoretically found observables, especially LC shapes, with some actual observed events and formulate conclusions based on these comparisons.

Not many detailed multidimensional simulations describing the aspherical interactions of SNe with their surroundings, consistently including RHD and non-local thermodynamic equilibrium (NLTE) radiation transfer (RT), have been performed to date \citep[see, e.g.,][and others]{1996ApJ...472..257B,vlasis16,2024ApJ...977..118O}. 
Our papers \citet[][hereafter \citetalias{PK19}]{PK19}; \citetalias{PK20}; \citet{Pejchaetall2022} provide certain insights into this problem using detailed 2D gas dynamic numerical simulations, while the latter performed 
calculations of observables semi-analytically, based on the determination of the shock-heating power per unit volume $\dot q$ from the adiabatic gas dynamic simulations and the subsequent conversion of the volume integrated $\dot Q =\int\dot q\,\text{d}V$ to bolometric luminosity $L_\text{SN}(t)$ using Arnett's law \citep[][see Eqs.~$8\,\text{\textendash}\,11$ in \citetalias{PK20}]{1980ApJ...237..541A,1982ApJ...253..785A}. 
In the current work, we fully employ 2D grey flux diffusion RHD, 
including radiative cooling using the CASTRO code \citep[e.g.,][]{2010ApJ...715.1221A,2011ascl.soft05010A}, as well as post-processed 2D Monte Carlo radiation transfer (MC-RT) calculations of observables 
using the codes SEDONA \citep{2006ApJ...651..366K,2010arXiv1008.2801A} and SIROCCO \citep{2025MNRAS.536..879M}.

The article is organised as follows: In Sect.~\ref{numsetup}, we describe the principles used to solve the RHD equations and the numerical solution using the MC-RT codes SEDONA and SIROCCO. Section~\ref{rhdset} presents the results of RHD models with stratified density levels. Section~\ref{sedlcs} demonstrates the calculated LCs, either pseudo-bolometric or in selected colour bands from three viewing angles. Section~\ref{sedspecpats} shows the calculated spectra as a function of the viewing angle, including the temporal evolution of selected lines. Section~\ref{sedpolsigs} demonstrates the semi-analytically calculated time evolution of relative polarisation for all models. In Sect.~\ref{discuss}, we discuss the results and their context, propose explanations for possible discrepancies, while Sect.~\ref{sumfutwork} summarises the article and outlines directions for further research in this area. In the Appendix, we describe the technical details 
and insert 
figures 
not included in the main article.

\section{Numerical setup}
\label{numsetup}
We describe, to a necessary extent, the equations of the RHD involved by CASTRO, as well as the equations of RT entering SEDONA and SIROCCO, 
including the 
initial and boundary conditions. 
\subsection{Description of RHD calculations}
\label{rhdcalc}
We solve the 2D RHD equations using the CASTRO grey radiation solver \citep{2011ApJS..196...20Z}, under the assumption of axial symmetry and the zero azimuthal velocity component of CSM (a negligible effect compared to the SN expansion). The fundamental equations in conservative form employed to solve the evolution of the RHD structure are
\begin{align}
&\frac{\partial\rho}{\partial t} + \vec\nabla\cdot\left(\rho\vec{u}\right) = 0,
\label{contione}\\[4pt]
&\frac{\partial\left(\rho\vec u\right)}{\partial t}+\vec\nabla\cdot\left(\rho\vec{u}\vec{u}\right)+\vec{\nabla}p+\lambda\vec{\nabla}E_r=0, 
\label{momentumone}\\[4pt]
&\frac{\partial E}
{\partial t}+\vec\nabla\cdot\left[\left(E+p\right)\vec u\right]+\lambda\vec u\cdot\vec\nabla E_r=-c\kappa_P
\left(aT^4-E^{(0)}_r\right),
\label{energyone}\\[4pt]
&\frac{\partial E_r}
{\partial t}+\vec\nabla\cdot\left(f^\prime\vec uE_r-\chi^\prime\nabla E_r\right)-\lambda\vec u\cdot\vec\nabla E_r=c\kappa_P
\left(aT^4-E^{(0)}_r\right),
\label{radone}
\end{align}
Hence, the continuity \eqref{contione}, momentum density \eqref{momentumone}, and total energy density \eqref{energyone} equations, respectively, with the grey radiation heating and cooling source term on the right-hand side of Eq.~\eqref{energyone}, are supplemented by the radiation energy density equation \eqref{radone}. In these equations, $\rho$, $\vec{u}$, $p$, $T$, and $E$ are the mass density, velocity, gas pressure, temperature, and total energy density. In radiative terms, $E_r$ is the radiation energy density, $\lambda$ is the flux limiter \citep{1981ApJ...248..321L}, $c$ is the speed of light, $a=4\sigma/c$ is the radiation constant ($\sigma$ is
the Stefan–Boltzmann constant), and $\kappa_P$ is 
the Planck mean opacity. The quantities $f^\prime$ and $\chi^\prime$ denote the terms $f^\prime=(3-f)/2$ and $\chi^\prime=c\lambda/\chi_R$, respectively, where $f$ is the Eddington factor and $\chi_R$ is the summed absorption and scattering mean Roseland opacity coefficient. The values of these coefficients and the relevance of their implementation are described further in Sect.~\ref{rhdboundcalc}. The terms with superscript $(0)$ ($E_r^{(0)}$) denote the corresponding quantity in the co-moving frame, while the terms without a superscript denote the quantity in the lab frame \citep[see][for details]{2010ApJ...715.1221A,2011ApJS..196...20Z}. The set of equations is closed with the ideal gas equation of state (EOS) $e=p/(\gamma-1)$ (gamma-law EOS in CASTRO), where $e$ is the gas internal energy density, along with the monatomic gas ratio of specific heats $\gamma=5/3$.

\subsection{Initial state of SN, spherical and aspherical CSM components}
\label{rhdinicalc}
Similar to \citetalias{PK20}, we first use the 1D SNEC code \citep{2015ApJ...814...63M} to propagate the SN shock wave through the progenitor, a red supergiant (RSG) star with a mass $M_\star\approx 12.3\,M_\odot$ and a radius $R_\star\approx 1000\,R_\odot$ at the moment of core collapse.
The explosion is initiated as a thermonuclear bomb with energy $10^{51}\,\text{erg}$, with $M=0.05\,M_\odot$ of radioactive nickel, and Paczy\'{n}ski EOS (radiation included). Regarding the higher masses of the CSM in the denser models, unlike \citetalias{PK20}, we now include the gravitational force
using the "Poisson's gravity" implicit solver in CASTRO; all other parameters are set to their default values. We do not consider the star's rotation due to its negligible effect on the investigated process. At the moment of shock breakout, we remap the following quantities from SNEC to CASTRO using its 1D spherically symmetric interpolation routine: density, velocity, temperature, pressure, and the abundances of H, He, C, O, Si, Fe, and Ni. However, in the hydrodynamic evolution using CASTRO, we no longer take radioactive heating into account.

We insert two components of the gaseous surroundings of an SN as an initial state: the spherically symmetric stellar wind and an aspherical component. 
Two types of aspherical CSM are implemented separately within five different models: two models of a circumstellar disc, labelled as DMIN and DMAX (hereafter DMs), differing by two orders of magnitude in initial density values; more parameters and initial scaling of the DMs models are described in Appendix~\ref{analmodels}. Three models of a bipolar lobe system are stratified by different initial density levels, formally labelled as models LMIN, LMOD, and LMAX (hereafter LMs). We list the basic parameters of all models in Tab.~\ref{tabmodels}: The first column containing numbers, $\dot{M}$, shows the progenitor's mass loss rate that produces the aspherical formation in the models. The second column $M_\text{w+d}$ quantifies the total mass of the surrounding CSM prior to the SN explosion; the third column $\rho_{0,\text{w}}$ lists the base densities of the spherically symmetric stellar CSM (wind), while the last column $\rho_{0,\text{ aspherical CSM}}$ shows the base densities in the circumstellar disc midplane in models DMIN and DMAX only because, in the case of bipolar lobes, such a base density cannot be identified.
\begin{table*}
\centering
\begin{threeparttable}
\caption{Basic input parameters of the models:\tnote{1}}
\label{tabmodels}
\setlength{\tabcolsep}{5.05pt}
\def\arraystretch{1.12}
\begin{tabular}{l|lcccc}
\hline
Type of aspherical CSM & Model & $\dot M\left(M_\odot\,\text{yr}^{-1}\right)$ & $M_\text{w+d}\left(M_\odot\right)$ & $\rho_{0,\text{w}}\,\left(\text{g}\,\text{cm}^{-3}\right)$ & $\rho_{0,\text{ aspherical CSM}}\,\left(\text{g}\,\text{cm}^{-3}\right)$\\\hline\noalign{\vskip 0.5mm} 
\multirow{2}{*}{Circumstellar disc} & DMIN & $\approx 10^{-10}$ & $\approx 4\times 10^{-1}$ & $6\times 10^{-16}$ & $5\times 10^{-13}$\\
 & DMAX & $\approx 10^{-8}$ & $\approx 2.4$ & $6\times 10^{-16}$ & $5\times 10^{-11}$\\\hline\noalign{\vskip 0.5mm}
& LMIN & $10^{-5}$ & $\approx 3\times 10^{-1}$ & $6\times 10^{-16}$ & $*$\\
Bipolar lobes & LMOD & $10^{-3}$ & $\approx 8\times 10^{-1}$ & $6\times 10^{-14}$ & $*$\\
& LMAX & $10^{-1}$ & $\approx 2.1$ & $6\times 10^{-12}$ & $*$\\
\hline
\end{tabular}
\begin{tablenotes}
\item[1] {\footnotesize Other parameters of individual models are listed in Appendix~\ref{analmodels}, further explanatory notes are in Sect~\ref{rhdinicalc}}.
\item[{\small *}] {\footnotesize Detailed parameters of bipolar lobes' complex structure are summarised in Tab.~\ref{tablobes}}.
\end{tablenotes}
\end{threeparttable}
\end{table*}

Both types of CSM models (disc vs. bipolar lobes) thus represent relatively large transfers of matter or outbursts in the pre-SN epoch. 
 We do not simulate an exact, observationally motivated model of the CSM, such as the Homunculus Nebula, in this work; we use its iconic structure only as a representation of a non-trivial morphology of lobes, forming transverse density barriers to the SN expansion and forcing SN-CSM interaction to be even more asymmetric \citep[see also other similar models in][]{2023eas..conf.1357K,2024eas..conf.1927K,2024IAUGA..32P2477K}. In the RHD model, as well as in RT calculations, we do not consider dust grains, which can be relevant for the evolution of complex plasma \citep{2023EPJD...77...56C} and for RT calculations, especially at longer infrared wavelengths. We will include dust in the future models of the SN-CSM interaction.   

\subsection{Boundary conditions, RHD code grid, and its setup} 
\label{rhdboundcalc}
The main RHD code parameters are as follows: we set the 2D cylindrical radial\textendash vertical ($\varpi$\textendash$z$) grid, where $\varpi\in \langle 0,R_\text{max}\rangle$, $z\in\langle -R_\text{max},R_\text{max}\rangle$, and $R_\text{max}=3.475\times 10^{16}\,\text{cm}\approx 500\,R_\star$ are defined with $4200\times 8400$ grid cells in the two specified directions, respectively; the numerical reconstruction scheme is the standard piecewise parabolic method (PPM). The boundary conditions (BC) for the gas-dynamic solution are as follows: symmetry BC at the left boundary (at $\varpi=0$, which represents the axis of the cylindrical system), while all the other three boundaries are outflow. The ideal-gas EOS, implementation of gravitation, and chemical elements are mentioned in Sects.~\ref{rhdcalc} and \ref{rhdinicalc}; we do not include a nuclear reaction network.

 Within the RHD solution, we employed the grey radiation solver stabilised by the flux limiter of the \citet{1981ApJ...248..321L} type, with the Dirichlet left BC, where $E_r=aT^4$ at $\varpi=0$, and Neumann BCs with the radiation flux $F_r=-\left(c\lambda/\kappa_R\right)\vec\nabla{E_r}$ at the three other boundaries.
 We parameterize, for simplicity, the initial constant Rosseland mean opacity $\kappa_R=0.34\,\text{cm}^{-1}$ (fit for electron scattering with the typical hydrogen abundance $X=0.7$), while the Planck mean opacity is not relevant here since the total opacity in the diffusion coefficient for the grey solver is set in CASTRO by the Rosseland mean opacity $\kappa_R$, and the Planck mean opacity is employed when the multi-group radiation solver is used \citep{2011ascl.soft05010A}. Incidentally, the value $\kappa_R=0.334$ from \citep[][]{web:TOPS:stats} corresponds to our situation at the moment of the SN shock breakout, when the values at the shock front from the SNEC calculation are $\rho\approx 3.12\times 10^{-9}\,\text{g}\,\text{cm}^{-3}$, $T\approx 2.9\times 10^5\,\text{K}$, and the relative abundances of key elements  are $0.67$ for H, $0.313$ for He, $0.0056$ for C, $0.0075$ for O, etc. For further parameters, see Appendix~\ref{analmodels}.

\subsection{Description of post-processed MC-RT calculations}
\label{MCRTdescr}
We use the data calculated in the RHD code as input for the MC-RT code SEDONA \citep{2006ApJ...651..366K,2010arXiv1008.2801A}, which is primarily designed for calculations of either LTE or NLTE radiation transfer in SNe. We do not use the simplified spherically symmetric hydrodynamic procedure in SEDONA with homologous expansion $v\propto r$; instead, we use individual output data files from the RHD code (a total of 250 time steps across all models) as a sequence of stationary MC-RT pre-calculations from which the SEDONA model file reads the geometrical parameters, density, temperature, radiative energy, velocity components, and the abundance of chemical elements involved (listed in Sect.~\ref{rhdinicalc}). The final emergent spectra and light curves of the SN model are obtained by collecting all escaping photon packets; they are binned in time of arrival, observed wavelength, and escape direction (i.e., viewing angle). A large number of bins in each dimension must be used to achieve the requisite resolution, along with enough packets collected in each bin to provide adequate photon statistics \citep{2006ApJ...651..366K}, outputting the $L_\nu(t_n,\mu)$ quantity, i.e., the monochromatic, directionally "observed" "luminosity".
  
\begin{figure*}[ht!]
 \centering
    \includegraphics[width=\textwidth]{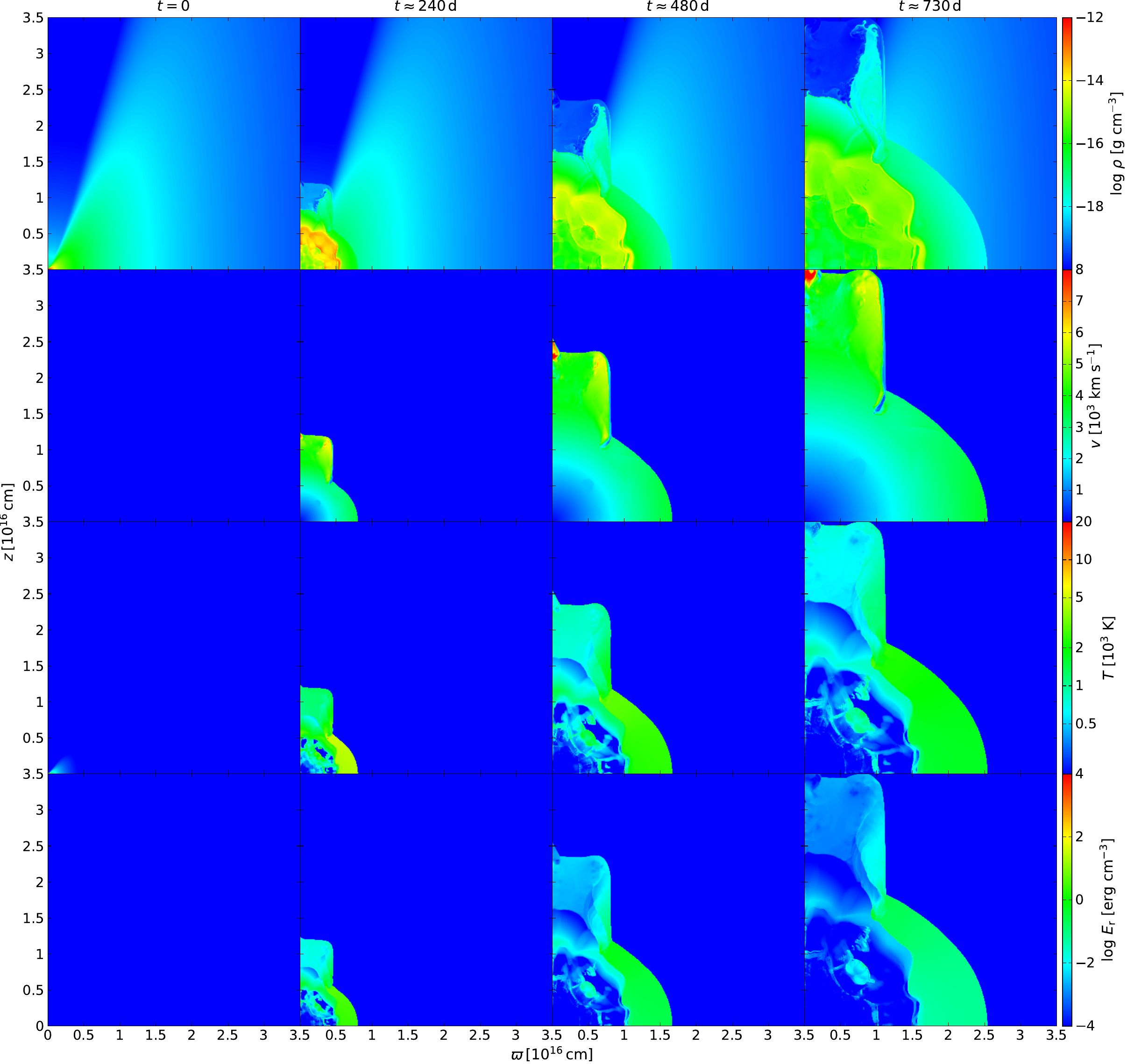}
    \caption{{\bf Model DMIN.} Selected evolutionary stages of SN interaction with a "thinner" circumstellar disk calculated by the RHD code. The four columns show the snapshots at times $t = 0$, $240$, $480$, and $730$~days. The individual rows shows four different quantities: from top to bottom it is density $\rho$, velocity magnitude $v$, temperature $T$, and radiative energy density $E_r$. The resolution of the bitmaps had to be significantly reduced  compared to resolution of the simulations in order to reduce the file size. For the animated version of this image, see Appendix~\ref{animasect}.}
\label{disk_first}
\end{figure*}
We use the frequency grid in the range $\nu_\text{min} = 10^{14} \,\text{Hz}$, $\nu_\text{max} = 2\times 10^{16} \,\text{Hz}$ for the LC calculations, corresponding to $150$\textendash$30\,000\,\AA$, logarithmically scaled with an increment $q=0.001$ (where $\text{d}\nu=q\nu$), resulting in 5300 frequency grid points, with $5\times 10^7$ photon packets ("particles") to initialise the simulations. We set 25 angular directions $\mu=\cos\theta$, where $\theta$ is the standard spherical polar angle, to cover a large range of observer's directions, from (nearly) polar to equatorial. We employ the NLTE regime for collisions and for calculations of the following opacities: electron scattering, bound-bound, free-free, and bound-free; for the elements listed in Sect.~\ref{rhdinicalc}, in variant configurations demonstrated in Fig.~\ref{spectra_elems}. Since we include the gas temperature from the RHD CASTRO calculations, we do not  update the gas temperature with opacity iterations; an additional test showed that whether we include this temperature update or not makes a negligible difference; moreover, a temperature update that includes only radiation would be inconsistent. We do not include the artificial SN core parameters. The bolometric (pseudo-bolometric, within the given frequency range) LCs (see Figs.~\ref{disk_lbolfirst} and \ref{lobes_lbolfirst}) and AB magnitudes \citep[][see Figs.~\ref{disk_third} and \ref{lobes_abmag}]{1965ARA&A...3...23O} for each model time step $n$ are calculated as
\begin{align}
\label{okolbol}
L_\text{bol}(t_n,\mu)=\int_{\nu_\text{min}}^{\nu_\text{max}}L_\nu(t_n,\mu)\,\text{d}\nu.
\end{align}
The AB magnitude for a frequency band [b] and an angular direction $\mu$ is 
\begin{align}
\label{okoabmag}
M_\text{AB}(t_n,\mu,\text{b})=-2.5\log_{10}\left[\frac{\int T_{\text{b}}(\nu)\,\nu^{-1}f_\nu(t_n,\mu)\,\text{d}\nu}{\int T_{\text{b}}(\nu)\,\nu^{-1}\,\text{d}\nu}\right]-48.60,
\end{align}
where $T_\text{b}(\nu)$ is the frequency-dependent transmission function for a given filter band [b]; the term $\nu^{-1}$ 
(in fact $(h\nu)^{-1}$, where the Planck constant $h$ cancels) 
indicates a photon energy-counting detector (CCD) response \citep[e.g.,][]{2007AJ....133..734B}, and the monochromatic spectral flux density $f_\nu=L_\nu/(4\pi d^2)$, where $d\equiv 10\,\text{pc}$. We note that within the numerically processed flux-filter convolution in Eq.~\eqref{okoabmag}, the tabulated frequency values of $T_\text{b}(\nu)$ for all selected bands must first be interpolated to the frequency grid of $f_\nu$ using the spline interpolation function.

We calculate the broadband spectra (see Fig.~\ref{spectra_elems}) in a shorter frequency range, covering the near-infrared, optical, and UV regions, roughly up to the Lyman jump, namely from $3.3\times 10^{14}\,\text{Hz}$ to $3.3\times 10^{15}\,\text{Hz}$, with 2305 frequency grid points and an increment of $q=0.001$ on a logarithmic scale. Unlike the LC calculation, we now involve the "last-iteration pump" function, where, in the last computational run, there are 10 times more photons initialised to achieve a higher final resolution of the spectral features. 

\begin{figure*}[ht!]
 \centering
    \includegraphics[width=\textwidth]{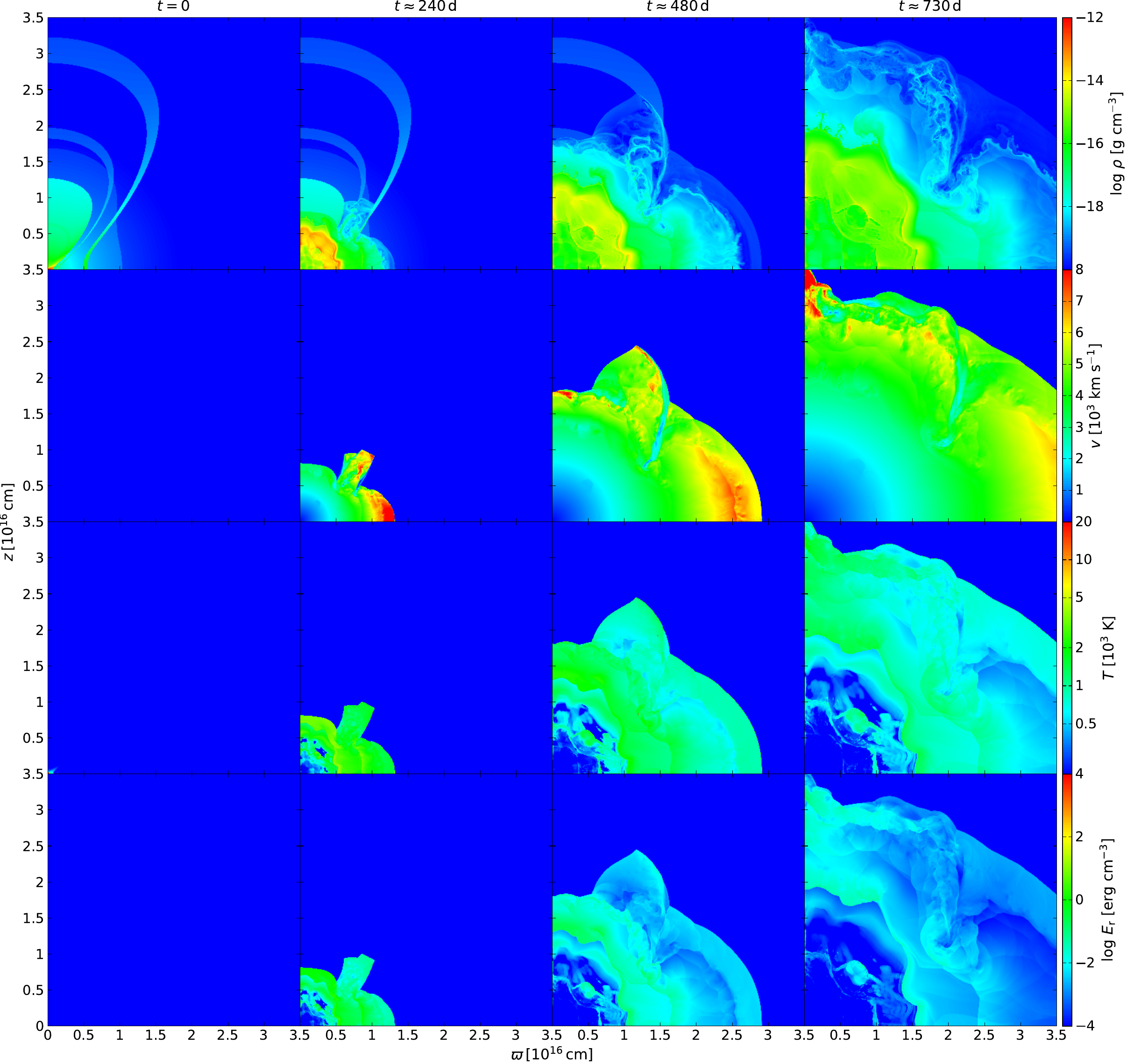}
    \caption{{\bf Model LMIN.} Selected evolutionary stages of the SN interaction with the "thinnest" bipolar lobes structure calculated by the RHD code. The four columns show the snapshots at times $t = 0$, $240$, $480$, and $730$~days. The individual rows show four different quantities: from top to bottom it is density $\rho$, velocity magnitude $v$, temperature $T$, and radiative energy density $E_r$. For the animated version of this image, see Appendix~\ref{animasect}.}
    \label{lobes_first}
\end{figure*}
We also demonstrate the time evolution of different line-of-sight profiles of a hydrogen spectral line, represented by  $\text{Ly-}\alpha$, in greater detail (see Fig.~\ref{spectra_lyman}); we calculate the individual line profiles using the Sobolev-based MC-RT code SIROCCO for 2.5-dimensional axisymmetric astrophysical outflows \citep{2025MNRAS.536..879M,2025ascl.soft10025L} in the "classic" mode (see Sect.~3.2 in \citet{2025MNRAS.536..879M}), where, unlike in the "hybrid macro-atom" mode, the wind itself may become a radiation source. We set the range $1150$\textendash$1280\,\text{\AA}$ with $10\,000$ frequency-grid points and $10^7$ photons per computational cycle. We inserted the mandatory central object with a mass $1.3\,M_\odot$, radius $7\times 10^{13}\,\text{cm}$, and a surface temperature $10^4\,\text{K}$ (roughly mimicking a neutron star remnant mass, surrounded by a relatively small optically thick environment), while the analytically imported parameters of a disc were excluded. The reason for choosing these parameters -- particularly the temperature of $10^4\,\text{K}$ -- in combination with the other parameters is that they approximate (albeit in a somewhat idealised manner) the actual conditions at a given distance from the centre of the simulation, and at the same time, best generate the $\text{Ly-}\alpha$ spectral line in the models at all the specified times. We note that the SIROCCO code requires the mandatory selection of a "central" object with a given mass, size, and temperature; however, its selected radius is so small (relative to the total extent of the computational domain) that it does not interfere significantly with the possible nebulosity of the entire configuration. For a similar calculation of the H$\alpha$ spectral line, a larger radius of $1.1\times 10^{14}\,\text{cm}$ was chosen in order to achieve a somewhat greater contribution to the total radiative flux, which proved to be essential for the clear generation of the H$\alpha$ line. 

We entered a one-component radiative wind of the imported type (whose parameters of coordinates, velocity components, density, and temperature are loaded from a file) in a cylindrical system.
The calculation of each spectral line was done in 25 ionisation cycles and 100 spectrum cycles (whose flowcharts are shown in Fig.~2 of \citet{2025MNRAS.536..879M}). The atomic data were imported from the SIROCCO "kurucz91.ls" library \citep{2025ascl.soft10025L}, containing $\simeq 10^4$ line transitions selected from the Kurucz CD-ROM 23 line-list with $\sim$$500\,000$ lines \citep{1995KurCD..23.....K}.

\section{RHD models of interaction}
\label{rhdset}
We have calculated the SN interactions with two morphologies of aspherical CSM: circumstellar disc and bipolar lobes (see \citetalias{PK20}), now with various levels of their pre-explosion density. 
We calculated our new RHD models 
using the CASTRO code \citep{2010ApJ...715.1221A,2011ascl.soft05010A}. In the main article, we show images of the RHD models with the lowest mass of aspherical CSM (DMIN and LMIN), as we consider them the most realistic; models with higher masses provide an interesting comparison and insight into the manifestations of SN-CSM interactions; however, given the observations, they are apparently mass-overestimated.
\subsection{SN interaction with circumstellar disc - DMs models}
\label{rhdsetdisk}
We have previously studied the interaction of SN ejecta with the circumstellar disc in \citetalias{PK19}; \citetalias{PK20} \citep[see also, e.g.,][and a few others]{vlasis16,2018ApJ...856...29M}, but only in an adiabatic hydrodynamic approximation, without including radiation hydrodynamics, which accounts for radiative cooling and heating.

Analogous to previous studies, we see that when the SN ejecta interact with the disc, the local shock wave propagates to some extent through the disc itself, but mainly into more "polar" regions outside the disc, generating an over-density shoulder near the vertical edge of the disc. In the {\bf DMIN model} (see Fig.~\ref{disk_first}), the ratio of the disc mass to the ejecta mass is relatively small (comparable to model A in \citetalias{PK20}), so the SN ejecta propagate relatively unimpeded into the disc region in the equatorial direction. The {\bf{DMAX model}} (see Fig.~\ref{disk_second}), with disc density parameters two orders of magnitude higher than in the DMIN model, shows (as expected) much greater deceleration of expansion in the disc direction, while the expansion of material in the polar direction is roughly similar to that in the DMIN model. However, there is a significantly greater over-density of a spiral-arm-like clump of accumulated material within the disc opening angle, which is not redirected in the polar direction. 

However, an important difference from the previous study appears to be the lower expansion velocity of the ejecta; for example, the shock region in the polar direction reaches an outer edge of the computational domain $R_\text{max}=450\,R_\star$ ($\approx 3.12\times 10^{16}\,\text{cm}$) in approximately 400\,days in the adiabatic solution in \citetalias{PK20}, while it reaches a slightly greater distance $\approx 3.5\times 10^{16}\,\text{cm}$ in approximately 700\,days within the RHD solution in the current study. Other hydrodynamic phenomena, such as the development of Kelvin-Helmholtz instabilities near the interface between freely expanding ejecta in the polar direction and the "edge" of the disc, remain essentially similar.

The left panel of Fig.~\ref{disk_lobes_helium_first} shows the density map of helium and iron in two more evolved times in model DMIN. We subsequently input the abundances of specific chemical elements (listed in Sect.~\ref{rhdinicalc}) into the code SEDONA for the calculation of broadband spectra, illustrating the intensity and time evolution of the corresponding spectral lines.

\subsection{SN interaction with bipolar lobes - LMs models}
\label{rhdsetlobes}
We also study the hydrodynamic evolution of SN ejecta colliding with nebulae consisting of bipolar lobes, resembling, for example, the structure of the well-known Homunculus Nebula surrounding the binary star $\eta\,\text{Car}$. However, its size is downscaled, and its total mass (and thus the density structure) is scaled in three levels, from lowest to highest: models LMIN, LMOD, and LMAX.
We can see that in the {\bf LMIN model}, the SN ejecta is only partially deformed according to the shape of the original
CSM; for example, in the equatorial direction, the ejecta can expand almost without slowing down, and its shock wave front in this direction is outside the computational domain at the 730-day ultimate simulation time. On the other hand, the densest lobe {\bf model LMAX} forces the ejecta to expand almost exclusively according to the shape of the original lobe structure, and the expansion velocity is significantly lower in this case. Furthermore, we can see that in the case of the thinnest LMIN model, the expansion velocity (in the initial and middle periods of the simulation) is highest in the equatorial direction; in the LMAX model, it is highest in the polar direction; and in the intermediate {\bf LMOD model}, it is highest in the intermediate direction of approximately $45^\circ$. Other hydrodynamic phenomena, except for the lower average expansion velocity (cf.~Sect.~\ref{rhdsetdisk}), correspond to those in the \citetalias{PK20} study, including reflecting shock waves and the fact that, due to the complex geometry of the CSM, it is difficult to identify various hydrodynamic
instabilities, as is the case in the simpler geometry of the circumstellar disc. Similar to the left panel of Fig. \ref{disk_lobes_helium_first}, and alongside other contexts for calculating observables, the right panel of Fig. \ref{disk_lobes_helium_first} illustrates the density map of helium and iron at two later times in model LMIN. 

\begin{figure*}[ht!]
    \centering
    \includegraphics[width=0.65\textwidth]{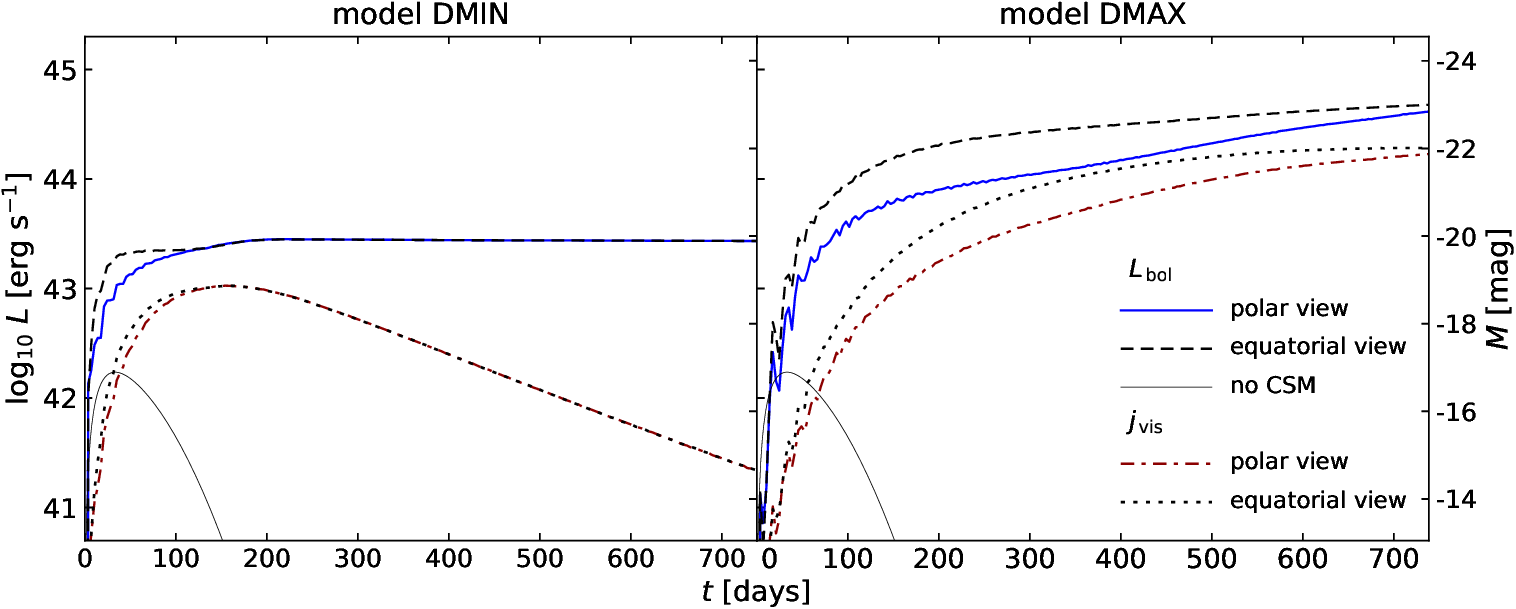}
    \caption{Pseudo-bolometric, directionally "observed" LCs of the two models of SN interaction with circumstellar disk-like structures, denoted as {\bf model DMIN} ({\it left panel}) and {\bf model DMAX} ({\it right panel}), up to approx.~2~years (see Sects.~\ref{rhdsetdisk} and \ref{SNdiskLCs}). The upper pair of lines shows the LCs calculated within the "largest" frequency range $10^{14}$\textendash$2\times 10^{16}$\,Hz (see Sect.~\ref{MCRTdescr}), while the bottom pair of lines shows the same within the range ($3.875$\textendash$7.825$)$\,\times 10^{14}$\,Hz, which roughly corresponds to the frequency range of visible light (for LCs corresponding to AB magnitudes in main filters, see Fig.~\ref{disk_third}). We demonstrate here the LC profiles for two different viewing angles in the graphs, the pole-on (polar) view marked in blue and the equatorial view marked in dashed black line for the bolometric LC (labelled $L_\text{bol}$ in the legend), and the polar view marked in red dash-dotted line and the equatorial view marked in dotted black line for the visual LC (labelled $j_\text{vis}$ in the legend). The black solid line labelled "no CSM" corresponds to an SN without CSM interaction, heated with $0.28\,M_\odot$ of radioactive Ni which roughly corresponds to the brightest normal H-rich SNe \citep{pejcha15,2017ApJ...841..127M} or the population mean of stripped SNe \citep{anderson19}.} 
    \label{disk_lbolfirst}
\end{figure*}
\begin{figure*}
 \centering
    \includegraphics[width=0.9\textwidth]
    {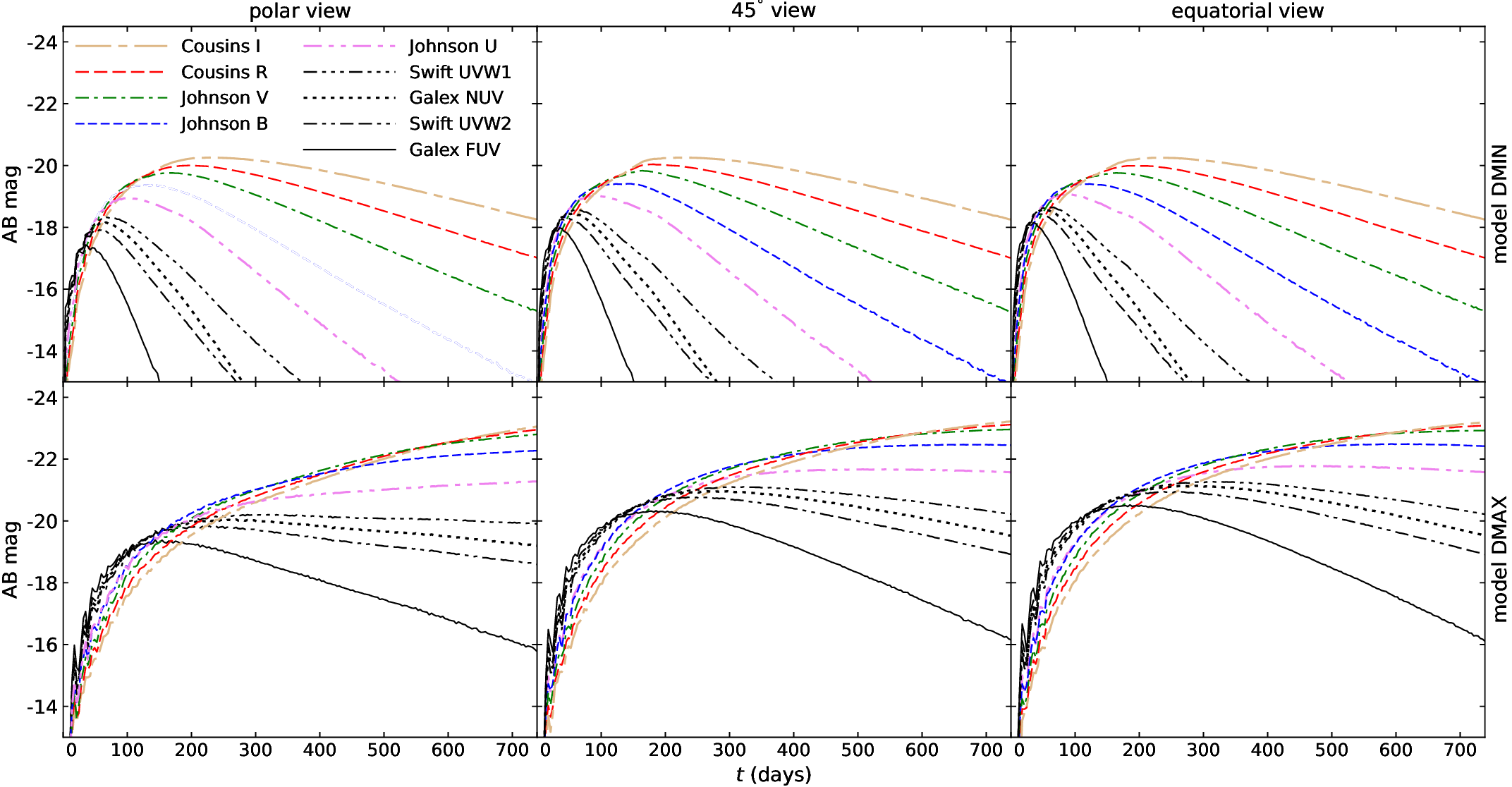}
    \caption{Calculated AB magnitudes for the two models of SN interaction with the circumstellar disk-like morphology. Each row in the figure corresponds to one of the models {\bf DMIN} or {\bf DMAX} labelled on the right. The three columns demonstrate the three various viewing angles marked at the top. The AB magnitudes are calculated for nine filters listed in the left top and bottom panels.}
    \label{disk_third}
\end{figure*}
\begin{figure*}
    \centering
    \includegraphics[width=0.9\textwidth]{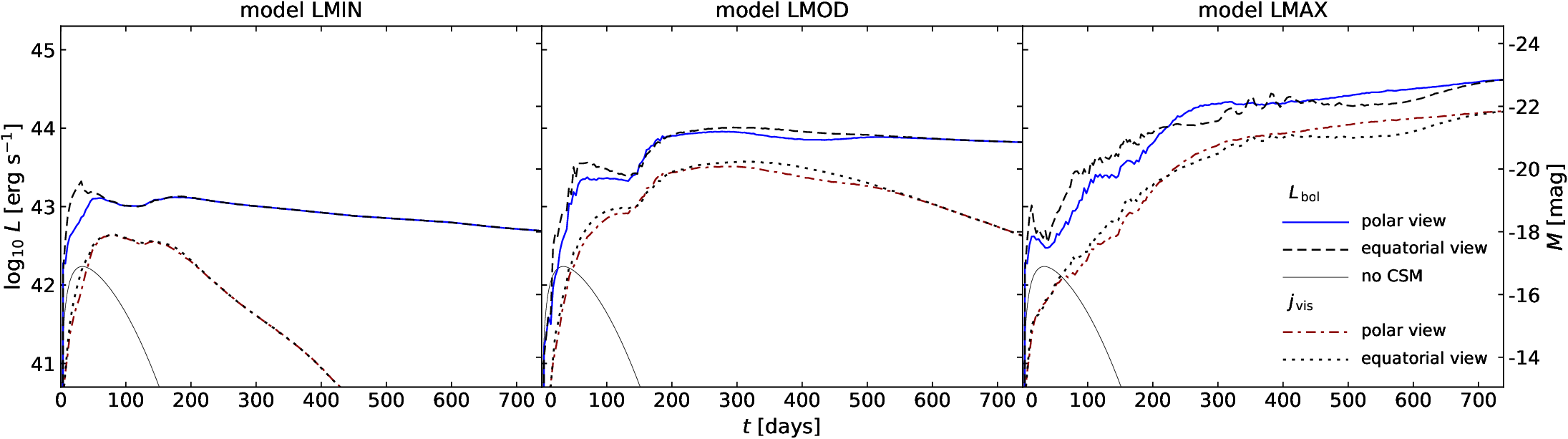}
    \caption{Pseudo-bolometric, directionally "observed" LCs of the three models of SN interaction with bipolar lobes structure, denoted as {\bf model LMIN} ({\it left panel}), {\bf model LMOD} ({\it middle panel}), and {\bf model LMAX} ({\it right panel}), up to approx.~2~years (see Sects.~\ref{rhdsetlobes} and \ref{SNlobesLCs}). The meaning and structure of the image, the order and types of lines, and the legend are the same as in Fig.~\ref{disk_lbolfirst}.} 
    \label{lobes_lbolfirst}
\end{figure*}
\begin{figure*}
 \centering
    \includegraphics[width=0.9\textwidth]
    {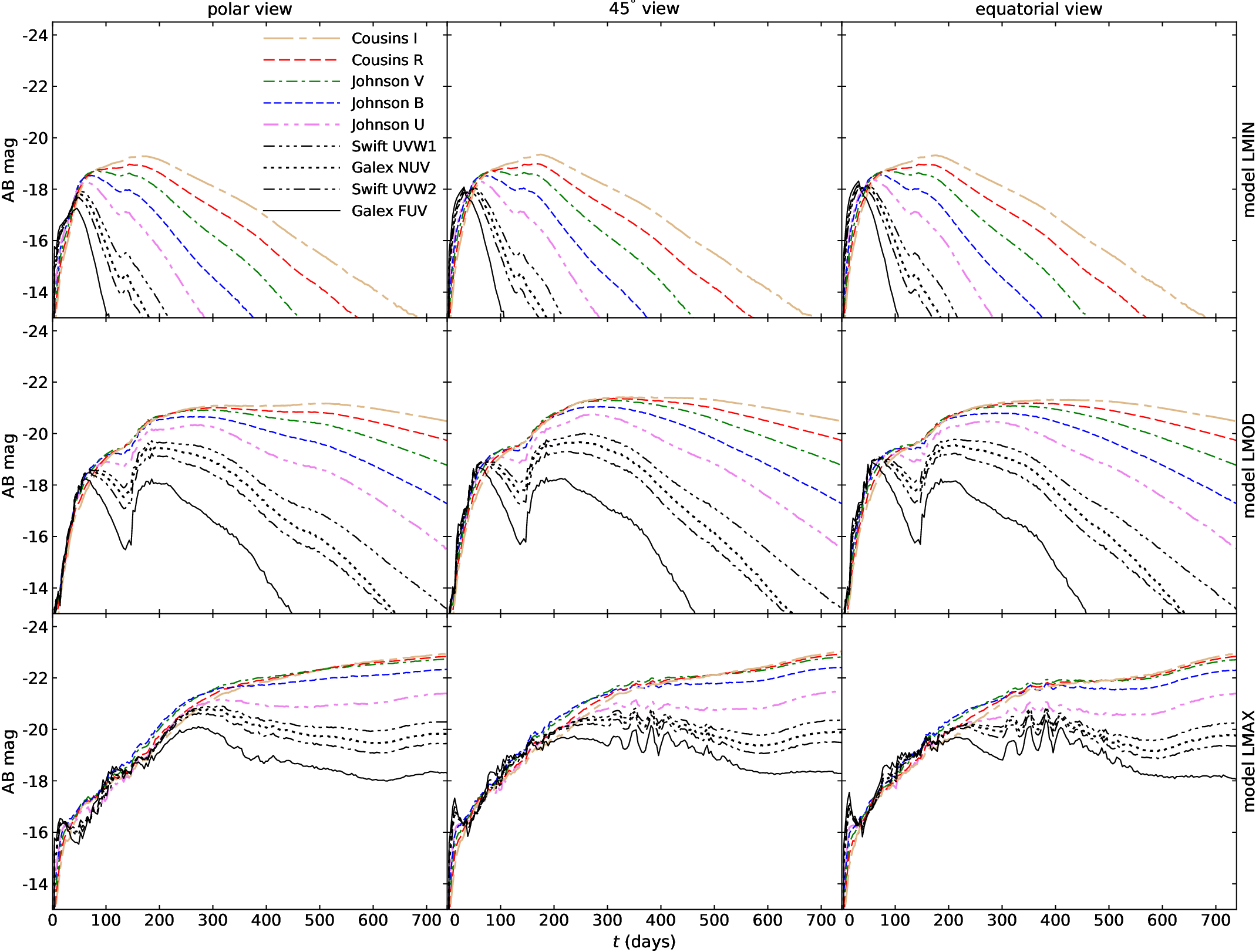}
    \caption{Calculated AB magnitudes for the three models of SN interaction with the bipolar lobes-like morphology. Each row in the figure corresponds to one of the models {\bf LMs}, labelled on the right. The other features of the figure are analogous to Fig.~\ref{disk_third}.}
    \label{lobes_abmag}
\end{figure*}
\section{Light curves and AB magnitudes}
\label{sedlcs}
Unlike in \citetalias{PK20}, where we calculated the LCs using the semi-analytical approach  \citep[Arnett's law,][]{1980ApJ...237..541A,1982ApJ...253..785A}, taking into account the semi-analytically calculated shock heating rate $\dot Q$, we used the 3D MC-RT code SEDONA \citep{2006ApJ...651..366K,2010arXiv1008.2801A} in this study to calculate robust observables (LCs, AB magnitudes, broadband spectra), while also having the ability to compare their appearance from different viewing directions.
\subsection{SN interaction with circumstellar disc - DMs models}
\label{SNdiskLCs}
Bolometric (pseudo-bolometric) as well as visual LCs calculated for models DMs (see Fig.~\ref{disk_lbolfirst}) show a relatively smooth evolution over time without significant drops, re-brightening, or bumps related to irregularities and asymmetries in the surrounding CSM, as well as the resulting changes in the energy balance, especially during the passage of shock waves in the SN ejecta as they pass through denser or rarer environments. This corresponds to the smooth nature of the surrounding CSM configuration and, in principle, also to the results of \citetalias{PK20}. In model DMIN, we see a significant difference in the polar and equatorial views only during the early phase, up to approximately 100 days. Equatorially "measured" LCs are even brighter at this time, which is probably due to the larger viewing angle at which the intense shock region with a higher temperature is unobscured by the optically thick region (cf. e.g., the third row's second panel in Fig.~\ref{disk_first}, even though this is a later time). This occurs regardless of the potentially higher shielding by the disc, as this shock area with high temperature is effectively shielded even in the polar direction to a greater distance by an expanded and, therefore, thicker layer of material ejected from the SN. Later, the two directional views show the same profiles due to the relatively rapid drop in the luminosity of shorter wavelengths, including UV, in this model (cf. Fig.~\ref{disk_third}), and the subsequent reprocessing of radiation into visible light, IR, and even longer wavelengths, where the re-emitted luminosity is practically the same in different directions. It is also evident that for up to approximately 50 days, the luminosity of this model is greater in shorter wavelengths, including UV, than in longer wavelengths.

For comparison, we have also included in Fig.~\ref{disk_lbolfirst} an analytically calculated LC (similar to that in \citetalias{PK20}, where we used Arnett's law, Eq.~10) of an SN powered by the decay of $0.28\,M_\odot$ of radioactive nickel (we remind that radioactive heating is not included in the current numerical models of SN-CSM interactions). In this context, it should be noted that the interaction energy that subsequently powers the LC is generated only in a certain section of the full spatial angle; therefore, the LC consists of two components: one generated by the interaction with dense, aspherical CSM and the other produced by non-interacting SN ejecta in the region outside the dense CSM. If, for example, the denser CSM were located further away from the progenitor star, there would be a double maximum in the LC: the first generated by the thermal energy of the SN explosion itself, plus the chain of radioactive decay. Only after the ejecta collide with the denser CSM would a secondary increase in luminosity caused by the SN-CSM interaction appear, as illustrated, for example, by the B series models in \citetalias{PK20}. In the theoretical models presented in this work, however, both of these components are combined into one because dense, aspherical CSM is located directly next to the progenitor star.

The LCs (bolometric as well as visual) are expected to be brighter in the DMAX model due to the greater interaction energy. The DMAX model also shows a larger difference between the LCs "observed" in the polar and equatorial directions, even in much later times. The reason for the higher luminosity in the equatorial direction is likely the same as that in the DMIN model. The reason for the much longer-lasting difference between the poles and the equator in terms of apparent luminosity is the much slower decrease in brightness at shorter wavelengths, where the difference is greater, as well as the increased brightness during the middle period, around 400\,\textendash\,500 days, in longer wavelengths, apparently due to a greater redistribution of radiative energy in denser CSM.

\subsection{SN interaction with bipolar lobes - LMs models}
\label{SNlobesLCs}
The bolometric and visual LCs in models LMs show a significant two-peaked profile with an intermediate drop at around 100 days. This "bumpy" evolution corresponds to the collisions of an SN shock expansion with relatively dense lobes positioned transversely to the direction of SN expansion propagation. The increase in luminosity as a result of the interaction is relatively steep from the outset due to the combination of the shock wave passing through the dense CSM in the polar direction and the inner lobes in the equatorial direction. The second luminosity peak is caused by the shock wave passing through the outer lobe in the equatorial direction. The later interaction of the SN with the lobes in the polar direction is manifested only by a slight undulation in the shape of the LC around 400\,\textendash\,500 days.

All other aspects are, in principle, similar to those in the case of interaction with the disc in Sect.~\ref{SNdiskLCs}. The model LMIN shows a sharp early maximum, particularly in the equator-ward luminosity (left panel in Fig.~\ref{lobes_lbolfirst}), caused by a dominant luminosity in short (UV) wavelengths, which is clearly visible from the top row of Fig.~\ref{lobes_abmag}. The model LMOD shows the most prominent yet relatively smooth drops in luminosity between SN-lobe interactions (see the middle panel in Fig.~\ref{lobes_lbolfirst} and the middle row in Fig.~\ref{lobes_abmag}). On the other hand, the model LMAX exhibits remarkable wave-like disturbances in shorter wavelengths and in the medium simulation time around 400 days, most likely caused by numerical instabilities in the MC calculation of radiative transfer in a relatively optically thick environment (see the right panel in Fig.~\ref{lobes_lbolfirst} and the bottom row in Fig.~\ref{lobes_abmag}).

The AB magnitude graphs in Fig.~\ref{lobes_abmag} clearly show much more pronounced peaks, rises, and falls in brightness at shorter wavelengths, especially in the UV region, which can thus perfectly trace any irregularities in the surrounding CSM and their distribution at these wavelengths. The detailed calculations of the AB and other types of magnitudes represent an important theoretical prerequisite for comparing luminosity, for example, in the UV range, particularly in connection with the planned UV missions, such as the wide-field survey mission ULTRASAT \citep[near-UV photometry;][]{2022SPIE12181E..05B,2024ApJ...964...74S}, the transient follow-up mission QUVIK \citep[Quick Ultra-VIolet Kilonovae surveyor;][]{2022SPIE12181E..0BW} for near-UV and far-UV high-cadence photometry, which is scientifically coordinated by the Masaryk University stellar and high-energy astrophysics teams \citep{2024SSRv..220...11W,2024SSRv..220...24K,2024SSRv..220...29Z,2025JATIS..11d2222Z}, and the planned UV photometry and spectroscopy mission -- Ultraviolet Explorer \citep[UVEX;][]{2021arXiv211115608K}.

\section{Spectral patterns}
\label{sedspecpats}
\subsection{Broadband spectral evolution}
\label{broadsedspecpats}
We calculate the time evolution of the broadband spectral features for the five specific models using the SEDONA code \citep[][see also the technical description of the code in Sect.~\ref{MCRTdescr} and the connection between the output from the CASTRO code and the SEDONA input therein]{2006ApJ...651..366K}. Figure~\ref{spectra_elems} demonstrates the broadband spectra in $3$~or~$4$ selected instant snapshots labelled in the pictures. Each spectral profile in each selected time is also represented by a split graph: coloured and black, showing either polar (coloured) or equatorial (black) viewing direction. 

For a better overview of the presence or absence of certain spectral lines, we show in Fig.~\ref{spectra_elems} these spectra divided into three columns, including a gradually increasing number of specific chemical elements: only H and He in the left column, medium-mass elements C, O, and Si in the middle column, and Fe and Ni in the right column. In the overall spectrum profile, especially in the case of heavier CSMs, jumps in the continuum are clearly visible, particularly the Balmer jump near $3650\,\mathrm{\AA}$ (the continuum curve is partially affected by the fact that the graphs are scaled in log-log form).

In this sense, the display of broadband spectra does not aim to show the detailed shape of spectral lines, as the current version of the SEDONA code does not include all the physics necessary for accurate nebular spectrum calculations (D. Kasen, private communication, October 10, 2025). For this reason, the SEDONA code apparently displays only the absorption curve of spectral lines without the corresponding emission. 

\subsection{Individual spectral-line evolution}
\label{detsedspecpats}
We also attach a brief study of the temporal evolution of a typical hydrogen spectral line, represented here by the Ly-$\alpha$ line. The technical details of the calculation process using the code SIROCCO \citep[e.g.,][also the private communication with S. Sim and Ch. Knigge, November 03 and December 10, 2025]{2025MNRAS.536..879M} and the justification for the choice of the Ly-$\alpha$ line are described in Sect.~\ref{MCRTdescr}. Figure~\ref{spectra_lyman} shows the evolution of this line in all five models, arranged in individual rows for three different viewing directions, represented by the three columns.

\begin{table*}
\caption{Values of the shape factor $\gamma$, averaged optical depth $\langle\tau\rangle$, and the 
maximum ($\theta=\pi/2$) relative polarisation $P_R$ for the models at four discrete times, 
$t= 100,\,300,\,500,\text{ and }730\,\text{d}$, calculated involving optically thin ($\tau\le 1$) regions of the expanding envelope (corresponding to right panel of Fig.~\ref{polari_lobes_lyman}).}
\label{tabpolar}
\begin{center}
\bgroup
\setlength{\tabcolsep}{5.05pt}
\def\arraystretch{1.12}
\begin{tabular}{l|ccr|ccr|ccr|ccr|}
\hline
\hline
Model & \multicolumn{3}{c|}{$t= 100\,\text{d}$}&
\multicolumn{3}{c|}{$t= 300\,\text{d}$}&\multicolumn{3}{c|}{$t= 500\,\text{d}$}&\multicolumn{3}{c|}{$t= 730\,\text{d}$}\\\cline{2-13}
   & $\gamma$  & $\langle\tau\rangle$ & $P_R$ (\%) 
   & $\gamma$  & $\langle\tau\rangle$ & $P_R$ (\%) 
   & $\gamma$  & $\langle\tau\rangle$ & $P_R$ (\%)
   & $\gamma$  & $\langle\tau\rangle$ & $P_R$ (\%)\\\hline DMIN & $0.9109$ & $0.8357$ & $-1.6200$ & $0.9413$ & $0.8667$ & $-1.7303$ & $0.9705$ 
& $0.8927$ & $-1.7152$ & $0.9777$ & $0.8657$ & $-1.6822$ \\
DMAX & $0.5460$ & $0.9845$ & $-0.6510$ & $0.7342$ & $0.9746$ & $-1.2417$ & $0.8771$ 
& $0.9725$ & $-1.7055$ & $0.9778$ & $0.9698$ & $-2.0507$ \\
LMIN & $0.8730$ & $0.8327$ & $-1.4318$ & $0.9277$ & $0.9015$ & $-1.5894$ & $0.9543$ 
& $0.8744$ & $-1.6242$ & $0.9674$ & $0.8713$ & $-1.6464$ \\
LMOD & $0.9321$ & $0.6584$ & $-1.1412$ & $0.9403$ & $0.8961$ & $-1.6202$ & $0.9217$ 
& $0.8942$ & $-1.6000$ & $0.9181$ & $0.8905$ & $-1.5644$\\
LMAX & $0.4196$ & $0.9761$ & $-0.7587$ & $0.4398$ & $0.9402$ & $-0.9022$ & $0.4533$ 
& $0.9913$ & $-0.9951$ & $0.8973$ & $0.9802$ & $-2.5738$ \\
\hline
\end{tabular}\vspace{0.35cm}
\egroup
\end{center}
\label{tabpopolar}
\end{table*} 
The Ly-$\alpha$ emission in the models is very characteristic, as it predominantly exhibits an asymmetric profile with multiple peaks. During the relatively early phase of up to approximately 100 days, Ly-$\alpha$ develops in the SN-disc (DMs) and SN-lobes (LMIN and LMOD) models from the polar view as a somewhat broad emission plateau, slightly shifted to the blue and concentrated in the range of approximately $-4000$ to $4000\,\text{km}\,\text{s}^{-1}$. In the intermediate view of $45^\circ$, such a plateau is clearly narrower for the given period, in the range of approximately $-2500$ to $2500\,\text{km}\,\text{s}^{-1}$, and is present only for SN-disc interactions, while for SN-lobes interactions, a more spiky profile prevails here. In the case of the equatorial view, this flat profile is very narrow to indistinguishable; especially for the SN-disc interaction model, it has a strikingly narrow central absorption corresponding to a dense, static environment in the direction of view. In the following period, after approximately 100 days, this "flattish" spectral line profile disappears and is replaced by a more or less double-peaked (horn-like) profile, which is very pronounced in the case of the SN-disc interaction \citep[cf., e.g., the theoretical spectral line profiles in][]{2017hsn..book..795J}. At the same time, it gradually transitions from slightly blue-shifted asymmetry to a red-shifted peak; we attribute this to the increasing rate of the resonant scattering of Ly-$\alpha$ photons
in neutral hydrogen moving outward in the SN ejecta or in the surrounding CSM (further discussion of this effect can be found in Sect.~\ref{discuss}).

Figure~\ref{spectra_balmer} demonstrates the typical, expected time-evolution of the correspondingly calculated H$\alpha$ spectral line. In the early (photospheric) phase, due to the relatively dense and extensive surrounding layers of the CSM lasting up to approximately 100 days, the H$\alpha$ emission profile exhibits only a very narrow emission peak, typical of Type IIn SNe, caused by the ionisation of the outer layers by the SN’s short-wavelength radiation. Later, after the transition from the photospheric to the nebular phase, the line broadens and forms a double-peaked profile due to emission from exposed, deeper, faster-moving layers, while the narrow emission peak produced by the ionised, static, or much slower-moving surrounding medium remains \citep[cf.,~e.g.,][etc.]{2015ApJ...814..108Y,2018MNRAS.475.1046I}. In addition, the H$\alpha$ profile, like the profiles of the other Balmer lines, is also predicted by these models to be much weaker (in terms of its equivalent width) than the Ly-$\alpha$. This is also indicated by Fig.~\ref{spectra_elems}, even though here it is for absorption.

The manifested time evolution of calculated spectral line details corresponds well to the measured profiles during an SN expansion. For example, \citet{smith15} demonstrates the case of measured H$\alpha$ spectral line evolution in the PTF11iqb SN, up to approximately 1100 days, with very similar evolutionary features in principle (see fig.~5 therein). The semi-analytical models A and C in \citetalias{PK20} also indicate the characteristic similarity in the basic properties of the evolution of a typical spectral line for the corresponding SN-CSM interaction morphologies.

\section{Polarisation signatures}
\label{sedpolsigs}
Similar to our \citetalias{PK20} study, we again estimate the degree of relative polarisation $P_R$ of the axially symmetric SN ejecta using the analytical formulas of \citet{1977A&A....57..141B} and \citet{1978A&A....68..415B}, including Thomson scattering in an optically thin environment irradiated by an optically thick internal source. Due to the complexity of calculating and estimating the position of the spatially fragmented photosphere within the expanding material, we do not calculate the polarisation numerically at this stage. The semi-analytical approximations mentioned above provide some insight into the morphologies of axisymmetric CSM and allow for a comparison of polarisation signatures depending on the density/mass of CSM in individual models.

The relative polarisation of Thomson scattering $P_R\,(\%)$ is given by the relation \citep[][Eq.~23]{1977A&A....57..141B}
\begin{align}
P_R\simeq\langle\tau\rangle\left(1-3\gamma\right)\sin^2\theta,
\label{relapol}
\end{align}
where the spherical polar angle $\theta=0$ indicates the polar viewing direction, and $\theta=\pi/2$ indicates the equatorial direction. We therefore provide the highest values according to the direction of observation (see Tab.~\ref{tabpolar} and Fig.~\ref{polari_lobes_lyman}), "seen" from the equatorial direction, while the values for other polar angles can be easily calculated using Eq.~\eqref{relapol}. The shape 
factor $\gamma$ and the averaged 
Thomson scattering optical depth $\langle\tau\rangle$ of the envelope \citep{1978A&A....68..415B} are
\begin{align}\label{polar2}
\gamma=\frac{\int_{r_\text{min}}^{\infty}\int_{-1}^{1}n\mu^2\,\text{d}r\,\text{d}\mu}
{\int_{r_\text{min}}^{\infty}
 \int_{-1}^{1}n\,\text{d}r\,\text{d}\mu},\quad\langle\tau\rangle=\frac{3}{16}\sigma_T
\int_{r_\text{min}}^{\infty}
\int_{-1}^{1} n\,\text{d}r\,\text{d}\mu,
\end{align}
where $\sigma_T$ is the Thomson scattering cross-section, $n(r,\mu)$ is the electron number density, and $\mu=\cos\,\theta$. The lower limit $r_\text{min}$ in radial integrals means either a fixed radius of $10^{16}\,\text{cm}$ or a limit of $\tau\le 1$ (for the two versions, see further in this paragraph). Assuming a completely ionised pure hydrogen envelope for simplicity, the electron number density is $n(r,\mu) = \rho(r,\mu)/m_H$, where $m_H$ is the mass of a hydrogen atom. Equation~\eqref{polar2} implies $\gamma = 1/3$ for a spherically symmetric morphology, while $\gamma < 1/3$ ($P_R > 0$) is for an oblate and $\gamma > 1/3$ ($P_R < 0$) for a prolate mass distribution.

Figure~\ref{polari_lobes_lyman} shows the relative polarisation $P_R$ calculation graph 
including 
the optically thin region $\tau\le 1$ for the 
polar angle $\theta=\pi/2$ viewing direction, calculated using the short characteristics solver \citep[][\citetalias{PK20}]{2018A&A...613A..75K}.
Since the formula \eqref{relapol} applies exclusively to optically thin media where $\tau\ll 1$, the estimate of the relative polarisation may be only marginally relevant; whereas including layers with even greater optical depth could yield non-physical results. However, since we consider this solution to be a starting point serving as a preliminary semi-analytical estimate for further, more advanced calculations, we include it here. 
The values of $\gamma$, $\langle\tau\rangle$, and $P_R$ for four selected times $t \approx 100,\,300,\,500$, and $730$ days are listed in Tab.~\ref{tabpopolar}. 
All models also show a prolate distribution of the ejecta with a relative polarisation $P_R<-1.5\%$ at later times (cf. fig.~13 in \citetalias{PK20}).
\begin{figure}
 \centering
    \includegraphics[width=\columnwidth]
    {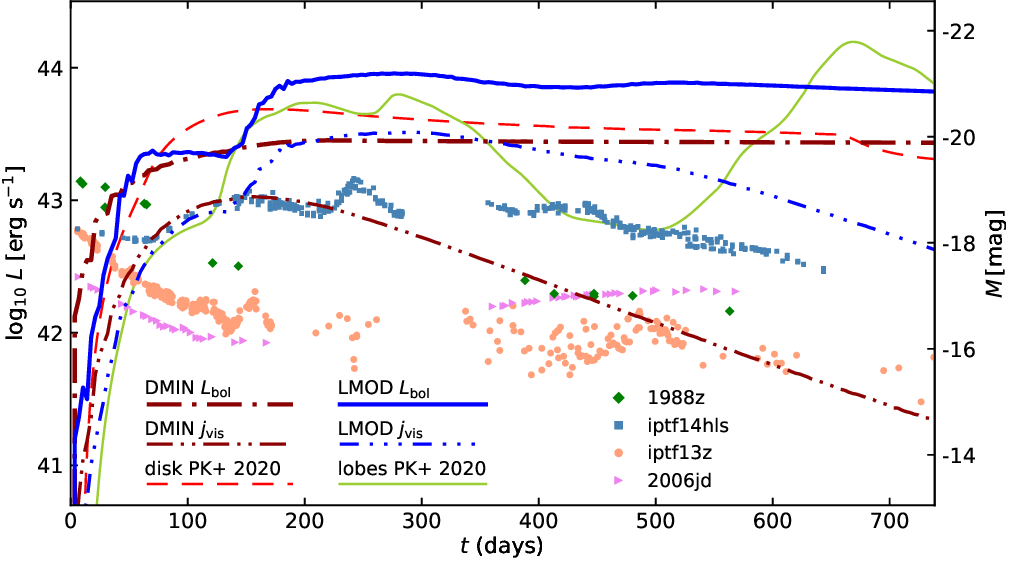}
    \caption{Bolometric and visual LCs of selected models. The thick blue and brown lines show the numerically calculated pole-on viewed LCs of the models DMIN and LMOD (cf. Figs.~\ref{disk_lbolfirst} and \ref{lobes_lbolfirst}). The red and green lines show the LCs of the SN-disk and SN-lobes interactions calculated semi-analytically in \citetalias{PK20}. The graph compares the synthetic LCs from our models calculated as a "first iteration" within the planned "chain" of models with the observed LCs of strongly interacting SNe of type IIn, SN1988Z \citep{aretxaga99}, iPTF14hls \citep{arcavi17}, iPTF13z \citep{2017A&A...605A...6N}, and 2006jd \citep{2012ApJ...756..173S}, shown with points.}
    \label{combLC}
\end{figure}
\section{Discussion}
\label{discuss}
\subsection{Relevance of the current models}
\label{discrelval}
The models shown here represent a systematic upgrade of some of our previous SN-CSM interaction models in \citetalias{PK20} in the following sense: we systematically performed more sophisticated simulations of RHD models, including a proper treatment of radiation in light-curves, broad-band spectral temporal evolution, and the time evolution of a particular angle-viewed spectral line profile to provide more realistic predictions for observations. The presented study does not currently aim to demonstrate exact compliance with specific SN-CSM interaction events; rather, it represents a "work iteration" of such a process in which it is challenging to find a configuration of the surrounding CSM in combination with the corresponding parametrisation of the progenitor star that would lead to observable characteristics fully consistent with some of the specific observed phenomena \citep[e.g.,][among others]{aretxaga99,2012ApJ...756..173S,arcavi17,2017A&A...605A...6N,2022MNRAS.516.1193K,2025arXiv251206067K,2026enap....2..720M}. At the same time, our models were primarily aimed at the technically sophisticated handling of all numerically conceived computational processes, including RHD modelling, MC-RT calculations of LCs in individual relevant bands, and spectra—both broadband and of selected spectral lines—all in a multidimensional form (currently as axisymmetric 2.5D in SIROCCO code), allowing their study depending on the observer's line of sight. This will allow us to estimate the situation for the next stage of the study, which will involve a better setting of such parameters and CSM configuration in order to achieve a more realistic match with the observed characteristics; for example, the initial decrease in the brightness of the LCs of interacting SNe, which is usually observed before irregular increases or re-brightenings occur, suggests that the dense CSM is probably located at a greater distance from the progenitor star. Only the calculation of relative polarizations has not been upgraded in this sense (see note in Sect.~\ref{sedpolsigs}), but its semi-analytical implementation is more detailed than in \citetalias{PK20} in terms of time evolution, as it is calculated for each time step and not just for a few selected moments.

Given the 3D SN--asymmetric disc-like CSM interaction simulations \citep[e.g.,][and a handful of others]{1996ApJ...472..257B,2024ApJ...977..118O}, we can anticipate differences or limitations inherent in current 2D models compared to potentially complete 3D models: Given that three dimensions imply more degrees of freedom, it is reasonable to expect a somewhat lower expansion velocity of the SN shock wave, as well as a more complex structure of instabilities, filaments, etc., particularly at the SN ejecta--CSM interface but also within the expanding envelope. For example, 3D simulations show that Rayleigh-Taylor (RT) instabilities develop more rapidly and are more pronounced than in 2D models. Emission spectra may also depend more variably on the viewing direction in 3D simulations; whereas 2D models show that emission from the SN-disc interaction is clearly influenced by the viewing angle, 3D simulations illustrate in greater detail how emission spectra can be blurred or swallowed up when the ejecta surrounds the disc. 

\subsection{Comparison with observed SNe}
\label{disccompar}
The LCs in the case of SN-disc interactions, calculated in our DMIN and DMAX models, are basically consistent with observations: both the bolometric and visual LCs show a relatively smooth and slow evolution over time; both curves in the DMIN model (left panel in Fig.~\ref{disk_lbolfirst} and Fig.~\ref{combLC}) reach maxima around $10^{43}$\,\textendash\,$10^{44}\,\text{erg}\,\text{s}^ {-1}$, while in the DMAX model (right panel in Fig.~\ref{disk_lbolfirst}), with a disc density two orders of magnitude higher, the LCs rise steadily over the entire 2-year period, with the maxima occurring roughly one order of magnitude higher. Given the values calculated from the current model, it does not appear that the combination with the contribution of radioactive decay (black solid line in Fig.~\ref{disk_lbolfirst}) significantly changes the resulting curve. The LCs of SN-lobe interactions (models LMs) show more or less distinct drops and increases. Figure~\ref{combLC} compares the calculated LCs, derived only from the DMIN and LMOD models for an easier overview, with some actual observed curves of prominent interacting SNe IIn, as well as with our previous semi-analytical calculations performed in \citetalias{PK20}. In particular, the LCs of these two models show a rough similarity to those determined semi-analytically in \citetalias{PK20}. Since the initial densities of the aspherical CSM were estimated somewhat ad hoc, they may be, for example, too high compared to actual situations. The purpose of this and subsequent studies is to gradually determine, based on comparisons, the parameters and morphology of the interacting environment that best correspond to the observed reality. 
For further considerations related to the course, evolution, and comparison of LCs with actually observed data, see Sects.~\ref{sedlcs} and \ref{discrelval}.

Within all models, we also calculate a synthetic profile of broad-band spectra, tracing the appearance or absence of spectral lines of a few selected important elements (see Sect.~\ref{broadsedspecpats} and the top two rows of Fig.~\ref{spectra_elems}) and a typical hydrogen spectral line from three different viewing polar angles (represented here by the Ly-$\alpha$ line; see Sect.~\ref{detsedspecpats} and the top two rows of Fig.~\ref{spectra_lyman}).
Few examples of observations of broad-band spectra include, e.g., \citet{bilinski18,bilinski20} in SN2013L and SN2014ab for H lines, \citet{mauerhan17} in SN2013ej with the evolution of Ca and O lines, or \citet{taddia20} in SN2013L with the evolution of Ca and Fe lines; all of these are in the visible and IR range. Among the most recent are, for example, \citet{2022MNRAS.516.1193K} for more H-rich SLSNe in the visible and near IR range, or \citet{2025arXiv251114916A} in the UV range of SN 2023taz, although the latter concerns hydrogen-poor SNe that show certain early evolutions of, for example, C, Mg, Ca, or Fe lines. The evolution of individual spectral lines (typically $\text{H}\alpha$) is documented, in addition to the articles already mentioned, for example, in  \citet{andrews18} for iPTF14hls, \citet{kokubo19} for KISS15s, \citet{szalai19,weil20} for SN2017eaw,
and \citet{andrews19} for SN2017gmr; see \citetalias{PK20} for a more comprehensive list. One of the important features of the spectral lines in this context is the double-peaked spectral profile associated with the disc or torus-like morphology of the surrounding CSM \citep[e.g.,][]{gerardy00,2017hsn..book..795J}. Such a double-peaked profile is evident in our DMs models, especially when viewed from the pole, and to a certain extent also in LMs; see Fig.~\ref{spectra_lyman}. The DMAX model with a dense disc also shows significant central absorption when viewed equatorially, i.e., a typical "shell" profile. The square profile (very approximate) with a flat peak and an indication of double-peaked evolution arises from bipolar lobes similar to those seen in $\eta$ Car. An example may be the eruption of the impostor SN UGC 2773-OT \citep{smith16}. Our models LMs in some evolutionary stages in Fig.~\ref{spectra_lyman} clearly resemble such profiles, with the exact shape of the spectral profiles depending on the viewing angle.

All models show a clear blue-red asymmetry of the spectral line in Fig.~\ref{spectra_lyman}, at almost all times and from all viewing directions. The transition from the initially slightly predominant blue peak to the later dominant red peak is striking. This phenomenon is well observed and documented, for example, in \citet{smith15} for PTF11iqb, including a discussion of possible connections with the presence of dust in the vicinity of an SN. Our current models indicate that this typical blue-red asymmetry in the Ly-$\alpha$ line should develop regardless of the dusty environment and irrespective of rotational asymmetry (which was an assumption discussed in \citetalias{PK20}). We note that no dust is included in the spectral calculations; furthermore, the dust, together with the expanding, relatively dense envelope, should tend to block the red part of the spectrum, as seen in many SNe \citep[cf.][]{smith15}. This developing asymmetry in the Ly-$\alpha$ line arises in the numerically calculated profile of spectral lines, even in completely axisymmetric models. The blue wing of the line may be stronger in the early phases because the hydrogen envelope has not yet expanded enough to develop a high column density or high-speed outflows; this allows "bluer" photons to escape directly rather than being scattered into the red wing. At later times, the photons undergo repeated resonant scattering by neutral hydrogen within the expanded nebula with higher optical depth; they diffuse in frequency to the red side of the line. The red-shifted photons "see" lower opacity, which allows them to escape more easily, creating a stronger red wing \citep[e.g.,][]{2012ApJ...751...29Y,2015ApJ...801L..16F}. On the other hand, the blue-red asymmetry of the H$\alpha$ spectral line, where the blue wing remains stronger than the red wing over an extended period, does not change significantly in Type II SNe, even in later stages; the emission originates from the inner regions of the ejected material, where the structure is predominantly formed by the initial explosion and remains relatively stable. The presence of the (however weak) line asymmetry indicates strong, ongoing interaction with a dense, asymmetric, H-rich CSM, providing direct evidence of extreme pre-SN mass loss \citep[e.g.,][etc.]{1996ApJ...472..257B,2025A&A...702A.213R}.

Our estimates of relative polarisation in the optically thin regime (see Tab.~\ref{tabpopolar} and 
Fig.~\ref{polari_lobes_lyman}) are essentially similar to those observed \citep[e.g.,][]{2011MNRAS.415.3497D,2019ARA&A..57..305G}. Aspherical CSM\,\textendash\,disks and bipolar lobes\,\textendash\,have a prolate ejecta shape and a maximum polarisation (in the equatorial line of sight) in the range of 0.5–3\% for all density levels. In \citetalias{PK20}, we determined the spherically symmetric radius of the He core as an optically thin boundary, while here we calculate the actual limit $\tau=1$ for the viewing direction $\theta=\pi/2$ (see the explanation in Sect.~\ref{sedpolsigs}).

\section{Summary and future work}
\label{sumfutwork}
We have upgraded the two-dimensional (2.5\,D) axisymmetric RHD simulations of SN-aspherical CSM interactions, incorporating a circumstellar disc and bipolar lobes with different mass levels.
The typical CSM masses within our grid are $0.3$ to $2.1\,M_\odot$ in the five presented models. Confrontation with observations will lead to corrections in the next stage of modelling. Snapshots of density, velocity magnitude, temperature, and radiative energy in the models DMIN and LMIN, which have the lowest masses, are shown in Figs.~\ref{disk_first} and \ref{lobes_first}; see also Sect.~\ref{rhdset}. The higher mass RHD models are presented in Figs.~\ref{disk_second}, \ref{lobes_first}, \ref{lobes_second}, and \ref{lobes_third}. Compared to purely adiabatic models in \citetalias{PK20}, current RHD models show greater deceleration of ejecta ($\sim$60\% of expansion velocity), while other features of gas-dynamics, including the deflection of ejecta material, the development of shock waves and instabilities, may be regarded as essentially similar.

Following our RHD simulations, the most significant upgrade consists of using the MC-RT code to calculate the following observable quantities of CSM interaction in SNe, depending on the viewing direction: LCs, both bolometric and visual, in selected filters, as well as broadband spectra in the range of $900 - 9000\,\AA$, with an evaluation of the presence of selected spectral lines, details of spectral line evolution (in this case Ly-$\alpha$), and relative polarisation (calculated semi-analytically). These calculated observable characteristics are described in Sects.~\ref{sedlcs}, \ref{sedspecpats}, and \ref{sedpolsigs}; demonstrated in Figs.~\ref{disk_lbolfirst}\,\textendash\,\ref{lobes_abmag} and \ref{spectra_elems}\,\textendash\,\ref{polari_lobes_lyman}. The observables show profiles and evolution that are generally consistent in the first "iteration" with the observations of Type IIn SNe, as well as with the semi-analytically calculated results of our previous studies \citetalias{PK19} and \citetalias{PK20}; see the plotted comparison of LCs in Fig.~\ref{combLC}. The evolution of the individual spectral line (Fig.~\ref{spectra_lyman}) shows the expected double-peaked profile in SN-disc interaction models and a flat-topped or multi-peaked profile in SN-lobe interaction models; however, even in the spherically symmetric configuration, a remarkable blue-red asymmetry evolves \citep[cf.,~e.g.,~the PTF11iqb H$\alpha$ line evolution in][or the 2018lqi and 2006gy H$\alpha$ line evolution in \citealt{2022MNRAS.516.1193K}]{smith15}.

In follow-up studies, we plan to further improve and refine the parametrisation and morphology of the SN progenitor and the surrounding CSM based on the agreement of observables with actual events. We also want to extend the computational domain to distances on the order of parsecs. In addition to more realistic luminous-blue-variable lobe scaling, we aim to calculate, for example, the interactions of SN with galactic discs, nuclear jets, and other structures in the vicinity of galactic nuclei \citep[cf.][]{2023Ap&SS.368....8C,2025MNRAS.540.1586K}. 

\begin{acknowledgements}
We thank Filip Hroch for his technical support and useful improvements in the computational process. We greatly appreciate the advice on the Castro code from Michael Zingale and Max Katz. We appreciate important discussions with Slah Abdellaoui and Jakub Podgorn\' y about polarisation properties. We also thank Dan Kasen for useful information about the code SEDONA, as well as Stuart Sim and Christian Knigge for their advice on the code SIROCCO. PK received support from the OPUS-LAP/GAČR-LA
bilateral project (2021/43/I/ST9/01352/OPUS22 and GF23-04053L). JKs and BK were supported by grant GA \v CR 25-15910S. MZ acknowledges the GA\v{C}R Junior Star grant no. GM24-10599M for support. Computational resources were supplied by the project "e-Infrastruktura CZ"
(e-INFRA LM2018140) provided within the programme Projects of Large Research,
Development, and Innovation Infrastructures.
\end{acknowledgements}

\bibliographystyle{aa} 
\bibliography{bibliography} 

\begin{appendix}
\onecolumn
\section{Additional graphs and figures}
\label{RHDsims}
\subsection{RHD simulation\texorpdfstring{\,\textendash\,}{ -- }model DMAX}
\label{RHDsimsDMAX}
\begin{figure*}[ht!]
 \centering
    \includegraphics[width=\textwidth]{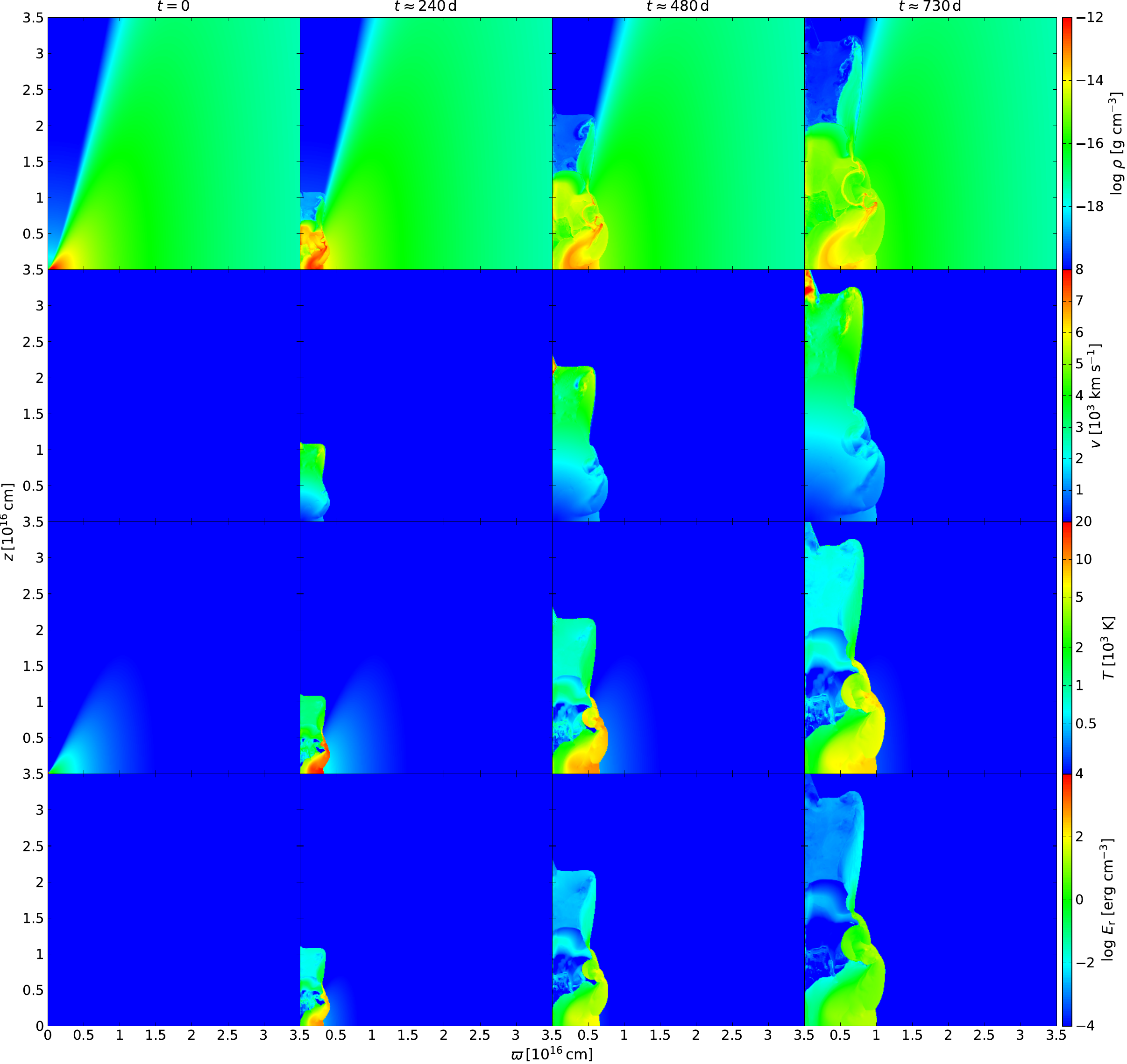}
    \caption{{\bf Model DMAX.} Selected evolutionary stages of SN interaction with a "denser" circumstellar disk structure calculated by the RHD code. The individual columns and rows show the same quantities as in Figs.~\ref{disk_first} and  \ref{lobes_first}. For the animated version of this image, see Appendix~\ref{animasect}.}
    \label{disk_second}
\end{figure*}

Figure \ref{disk_second} shows the RHD evolution snapshots of the DMAX model using the same structure as in Fig.~\ref{disk_first}; see also the description in Sect.~\ref{rhdsetdisk}. In this model, among other things, it is clearly visible how the dense inner part of the disc is pushed away by the expansion of the SN, creating a gradually thinning, propeller-shaped structure that roughly fills the angular range of the disc's opening angle. Unlike the DMIN model (remember that the initial density of the aspherical component CSM – the disc – is two orders of magnitude higher than in the DMIN model), the expansion of the SN ejecta in the equatorial direction is almost completely blocked. The reason for this is the rapid loss of momentum in the direction of the disc and its deflection into areas of lower density. This evolution is also reflected analogously in the temperature structure and the structure of radiative energy, while the development of the velocity structure shows the opposite effect. 
\subsection{RHD simulation\texorpdfstring{\,\textendash\,}{ -- }model LMOD}
\label{RHDsimsLMOD}
\begin{figure*}[ht!]
 \centering
\includegraphics[width=\textwidth]{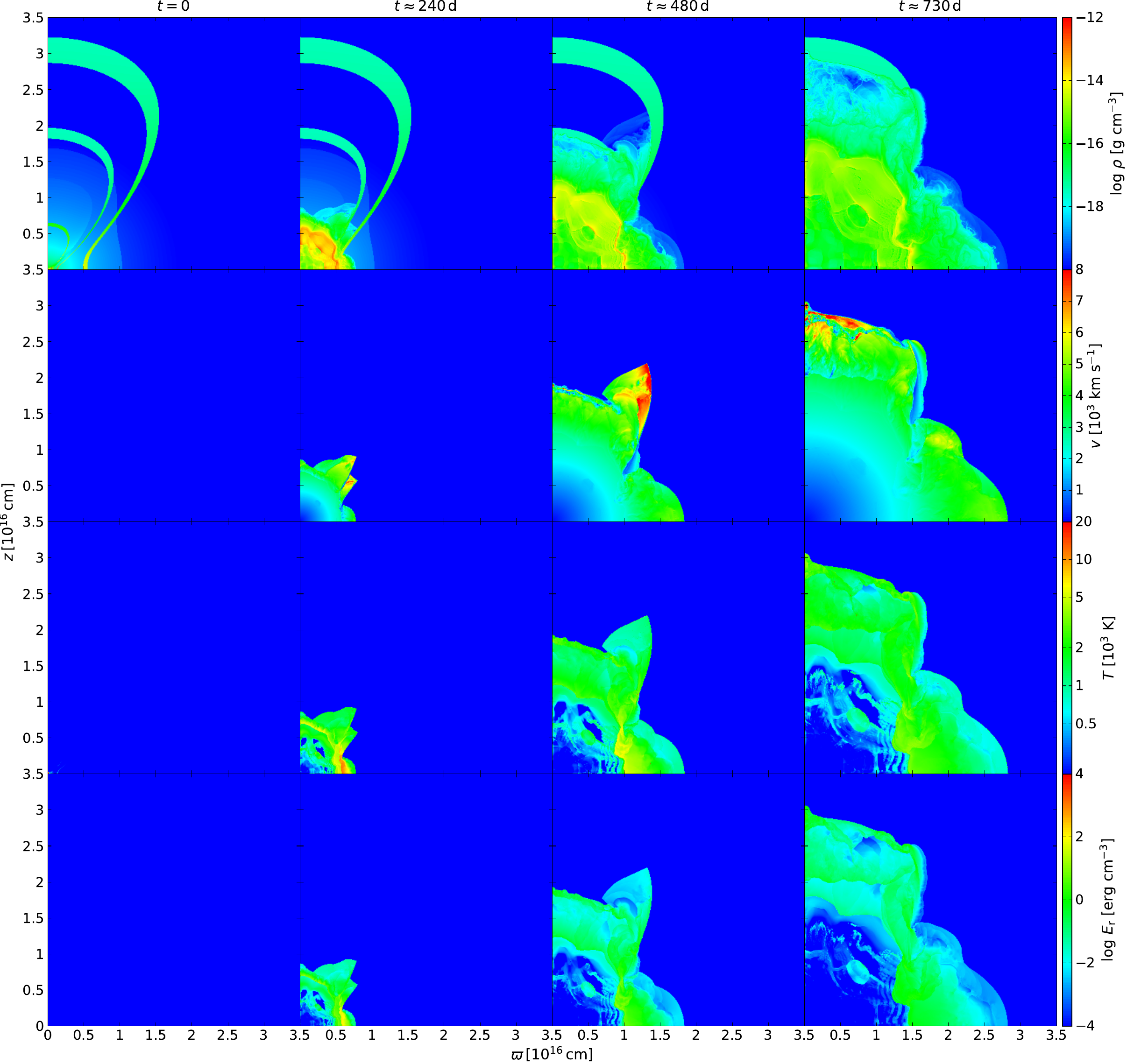}
    \caption{{\bf Model LMOD.} Selected stages of SN interaction with a "moderate density" bipolar lobes structure calculated by the RHD code. The individual columns and rows show the same quantities as in Figs.~\ref{disk_first}, \ref{lobes_first}, or \ref{disk_second}. For the animated version of this image, see Appendix~\ref{animasect}.}
    \label{lobes_second}
\end{figure*}

Figure \ref{lobes_second} shows the RHD evolution snapshots of the LMOD model, using the same structure as in Fig.~\ref{lobes_first}; see also the description in Sect.~\ref{rhdsetlobes}. Unlike the LMIN model (remember that the initial density of the aspherical component CSM – the bipolar lobes – is two orders of magnitude higher on average than that in the LMIN model), where the maximum expansion velocity heads towards the equatorial direction, in the LMOD model, the expansion of the SN ejecta in both the equatorial and polar directions is significantly decelerated by the denser structure of the transversely positioned lobe slabs, especially by the much higher mass-concentrated narrow waist. As a result, the highest velocity magnitude appears towards the mid-positioned $45^\circ$ angular direction, forming a significant protrusion arm there. 
\subsection{RHD simulation\texorpdfstring{\,}{ }\textendash\texorpdfstring{\,}{ }model LMAX}
\label{RHDsimsLMAX}
\begin{figure*}[ht!]
 \centering
    \includegraphics[width=\textwidth]{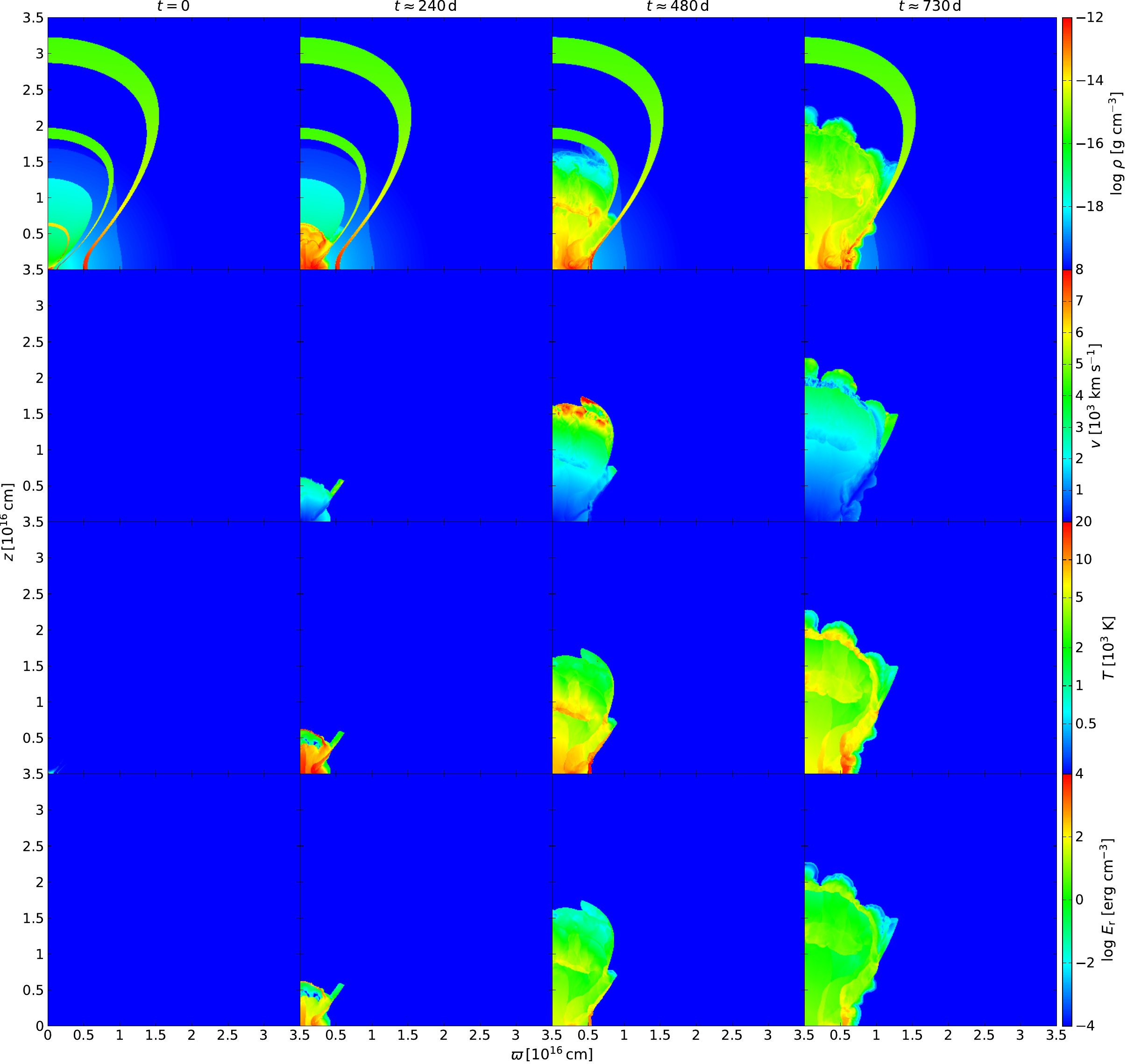}
    \caption{{\bf Model LMAX.} Selected stages of SN interaction with a "densest" bipolar lobes structure calculated by the RHD code. The individual columns and rows show the same quantities as in Figs.~\ref{disk_first}, \ref{lobes_first}, \ref{disk_second}, or \ref{lobes_second}. For the animated version of this image, see Appendix~\ref{animasect}.}
    \label{lobes_third}
\end{figure*}

Figure \ref{lobes_third} shows the RHD evolution snapshots of the LMAX model using the same structure as in Fig.~\ref{lobes_first}; see also the description in Sect.~\ref{rhdsetlobes}. Since the initial density of the aspherical component CSM – the bipolar lobes – is two orders of magnitude higher than that in the LMOD model on average (four orders of magnitude higher than in the LMIN model), the SN expansion is almost entirely blocked by the lobe shells; this effect practically does not allow the central parts of the expanding SN configuration to rarefy. However, the relative expansion velocity is, in this case, highest towards the pole, where the efficiently deflected mass causes the highest concentration of momentum in this direction. The expansion of the SN ejecta in the equatorial and intermediate directions is almost entirely blocked. The expanding envelope is effectively slowed down by dense lobes where it accumulates for some time, thus emphasising their original structure, which is particularly visible in plots of density, temperature, and radiative energy.

In addition, Fig.~\ref{disk_lobes_helium_first} shows the evolution of He and Fe densities in the models DMIN and LMIN at selected time snapshots. The structure of the plots is described in Sects.~\ref{rhdsetdisk} and \ref{rhdsetlobes}.
\clearpage
\subsection{MC-RT simulations}
\label{appmontecarsims}
\begin{figure*}[ht!]
 \centering
    \includegraphics[width=0.97\textwidth]{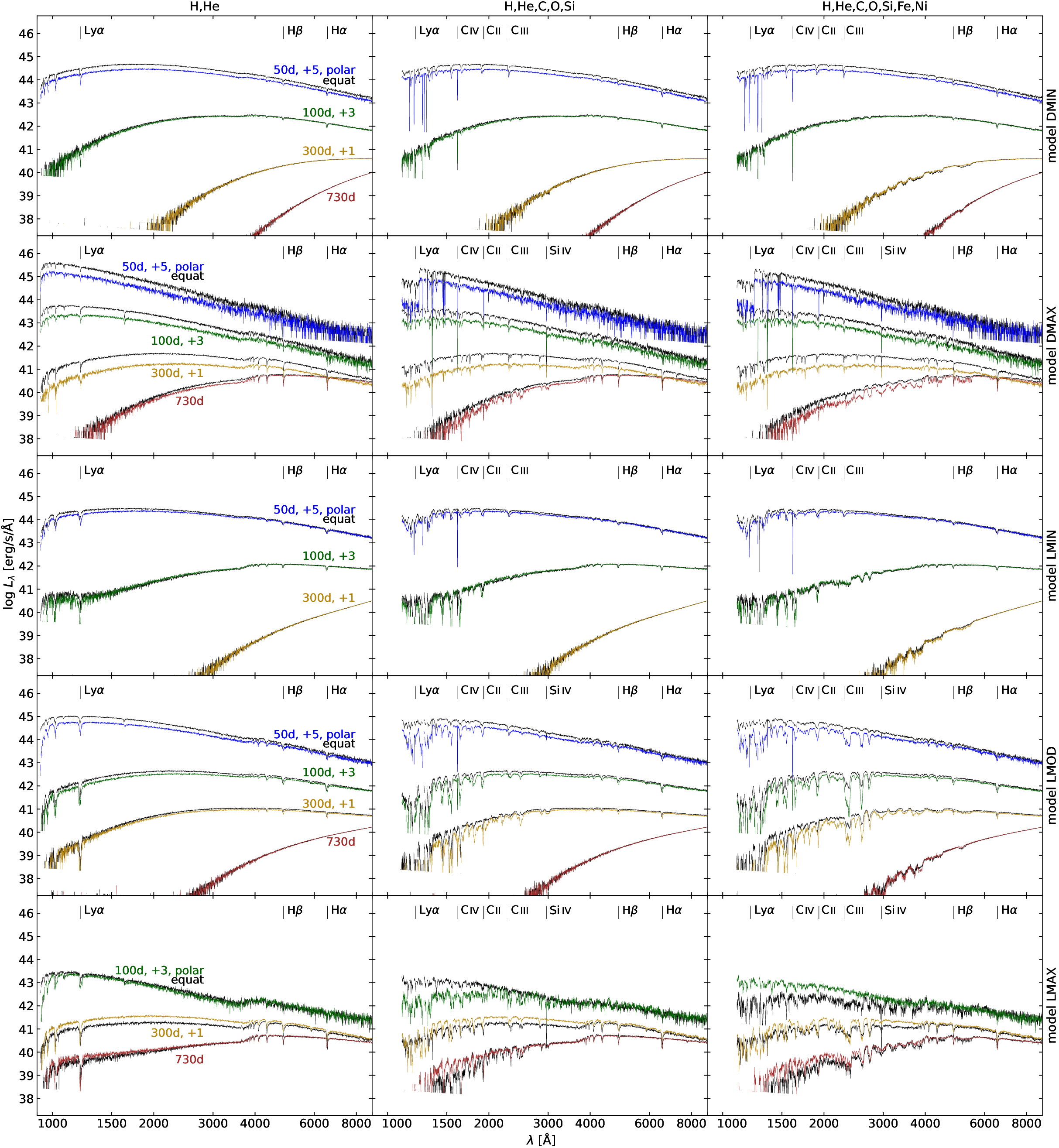}
    \caption{The overall behaviour of the spectral features calculated for all models marked on the right for individual rows, using the MC-RT code SEDONA \citep{2006ApJ...651..366K} under conditions described in Sects.~\ref{numsetup} and \ref{sedspecpats}. The three columns enhance the difference between the spectra if only H and He are included ({\it left column}), some intermediate mass elements C, O, and Si are added ({\it middle column}), and Fe with Ni are added ({\it right column}). The times of particular spectra snapshots are marked directly in the image next to the corresponding graph. The coloured graphs show the spectra calculated from the direction of the pole, while the paired black graphs are calculated from an equatorial perspective. For clarity, certain pairs of spectra are shifted upwards by the positive values shown next to them. Some significant spectral lines are marked in the graph. Blends of iron lines in the right column are not marked.}
    \label{spectra_elems}
\end{figure*}

\begin{figure*}[ht!]
 \centering
    \includegraphics[width=0.75\textwidth]{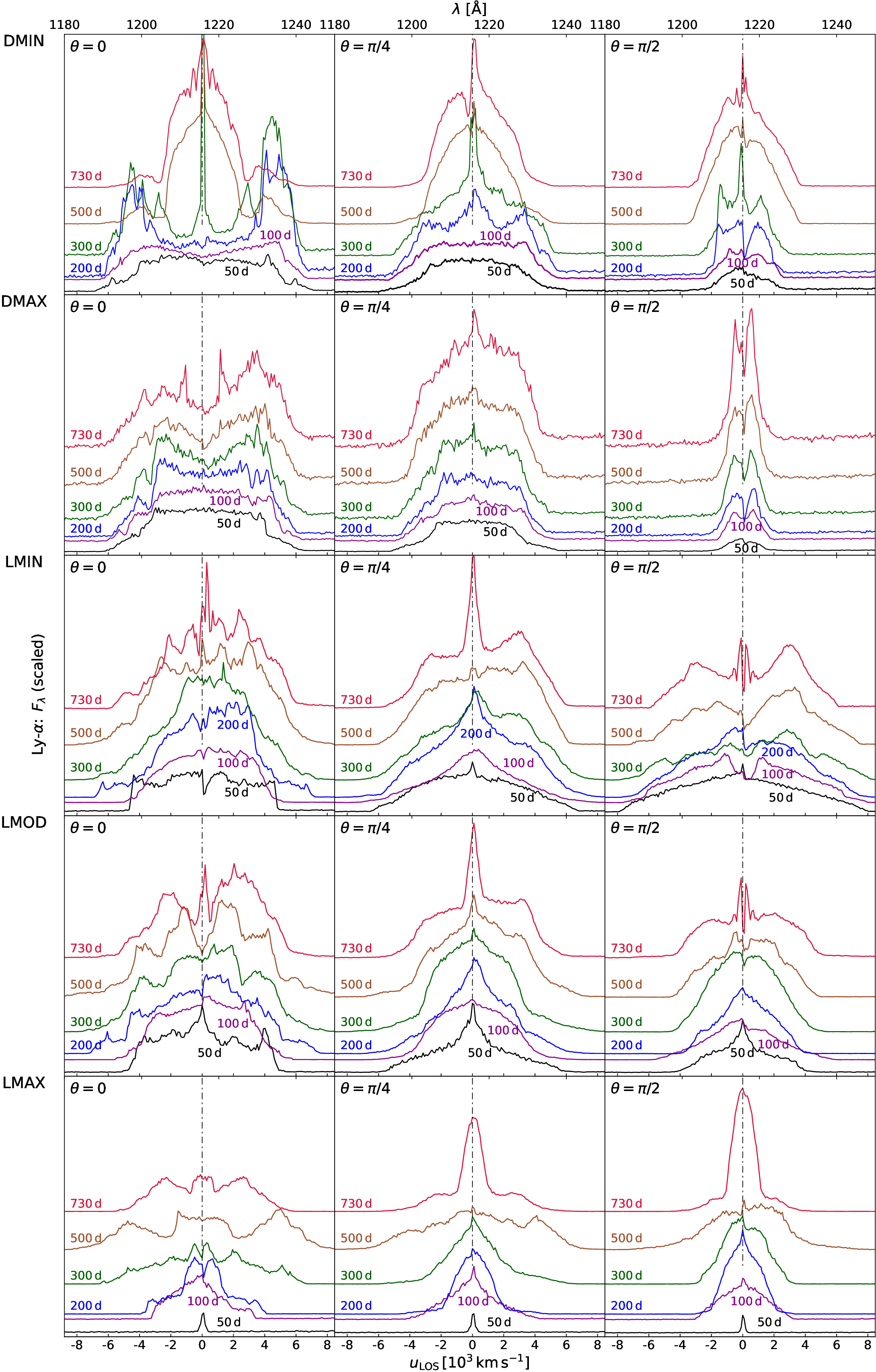}
    \caption{The time evolution of a hydrogen spectral line. In this case, the detailed calculations of Ly-$\alpha$ line were made using the MC-RT code SIROCCO \citep{2025MNRAS.536..879M} under conditions described in Sects.~\ref{numsetup} and \ref{sedspecpats}. The individual rows show a particular model, labelled on the left; each column corresponds to a different viewing polar angle $\theta$ denoted within each panel. The times of shown simulations range from 50 to 730 days with the particular time-snapshots emphasized by different colours and labelled directly in the plot. Spectral line profiles are plotted on a linear scale and normalised for clarity of presentation; for this reason, we do not display units on the vertical axis (cf.~fig.~12 in \citetalias{PK20}). The originally slightly blue-shifted peak of emission transfers after about 300\,\textendash\,500 days to a red-shifted peak, as corresponds to observations of the spectral lines of SNe \citep[see, e.g.,][]{smith15}.}
    \label{spectra_lyman}
\end{figure*}

Figures \ref{spectra_elems} and \ref{spectra_lyman} show the temporal evolution of broadband spectra and the individual spectral line Ly-$\alpha$  of the presented models, respectively, calculated numerically using the MC-RT codes. The structure of these images and the nature of the physics behind them are sufficiently described in Sects.~\ref{broadsedspecpats} and \ref{detsedspecpats}. In addition, Fig.~\ref{polari_lobes_lyman} shows the semi-analytically calculated relative polarisation evolution; see Sect.~\ref{sedpolsigs} for further description and explanation.
\begin{figure*}[ht!]
 \centering
    \includegraphics[width=0.75\textwidth]{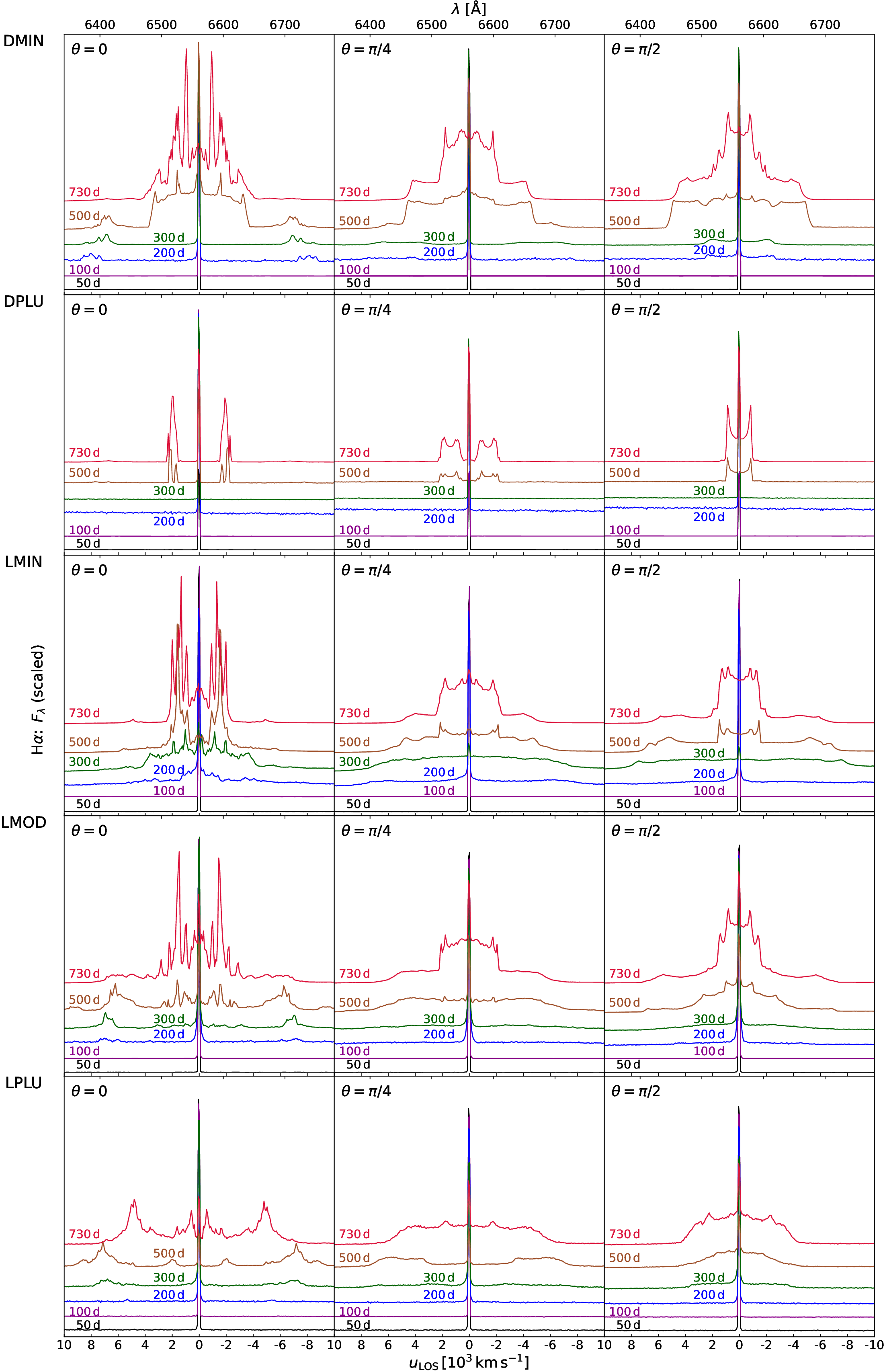}
    \caption{As in Fig.~\ref{spectra_lyman}, now the detailed calculations of H$\alpha$ line. The characteristic central emission peak is significantly narrow while the double-peaked profile begins to appear only at later time, when the process enters the nebular phase (remarkable after some 200-300 days);
    the earlier profiles exhibit, except the central narrow peak, only an irregular noise within a continuum. See the further description in Sect.~\ref{detsedspecpats}.}
    \label{spectra_balmer}
\end{figure*}
\begin{figure*}[ht!]
 \centering
    \includegraphics[width=0.54\textwidth]{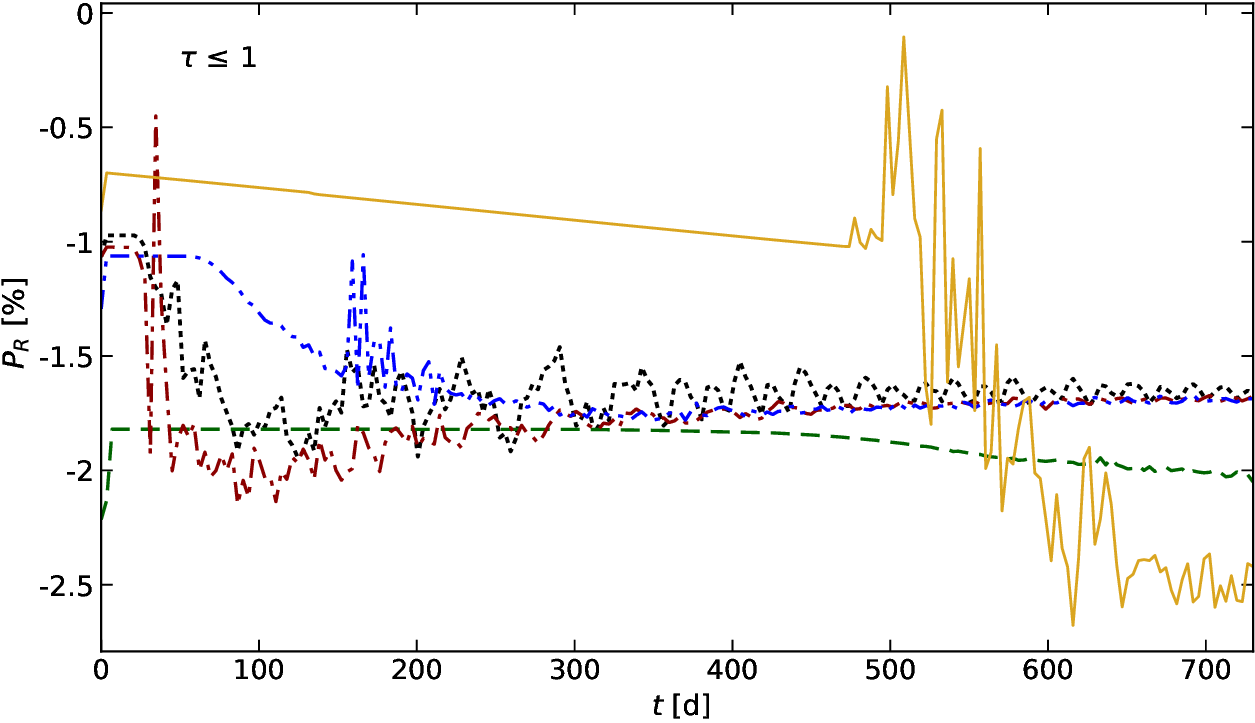}
    \caption{Graph of maximum (for $\theta=\pi/2$) relative polarisation evolution for the five models, including 
    only the optically thin region ($\tau\le 1$). 
    The values 
    at selected times are listed in Tab.~\ref{tabpolar}. (See also the polarisation calculations for models in \citetalias{PK20}.)}
    \label{polari_lobes_lyman}
\end{figure*}
\begin{figure*}[ht!]
 \centering
    \includegraphics[width=0.45\textwidth]{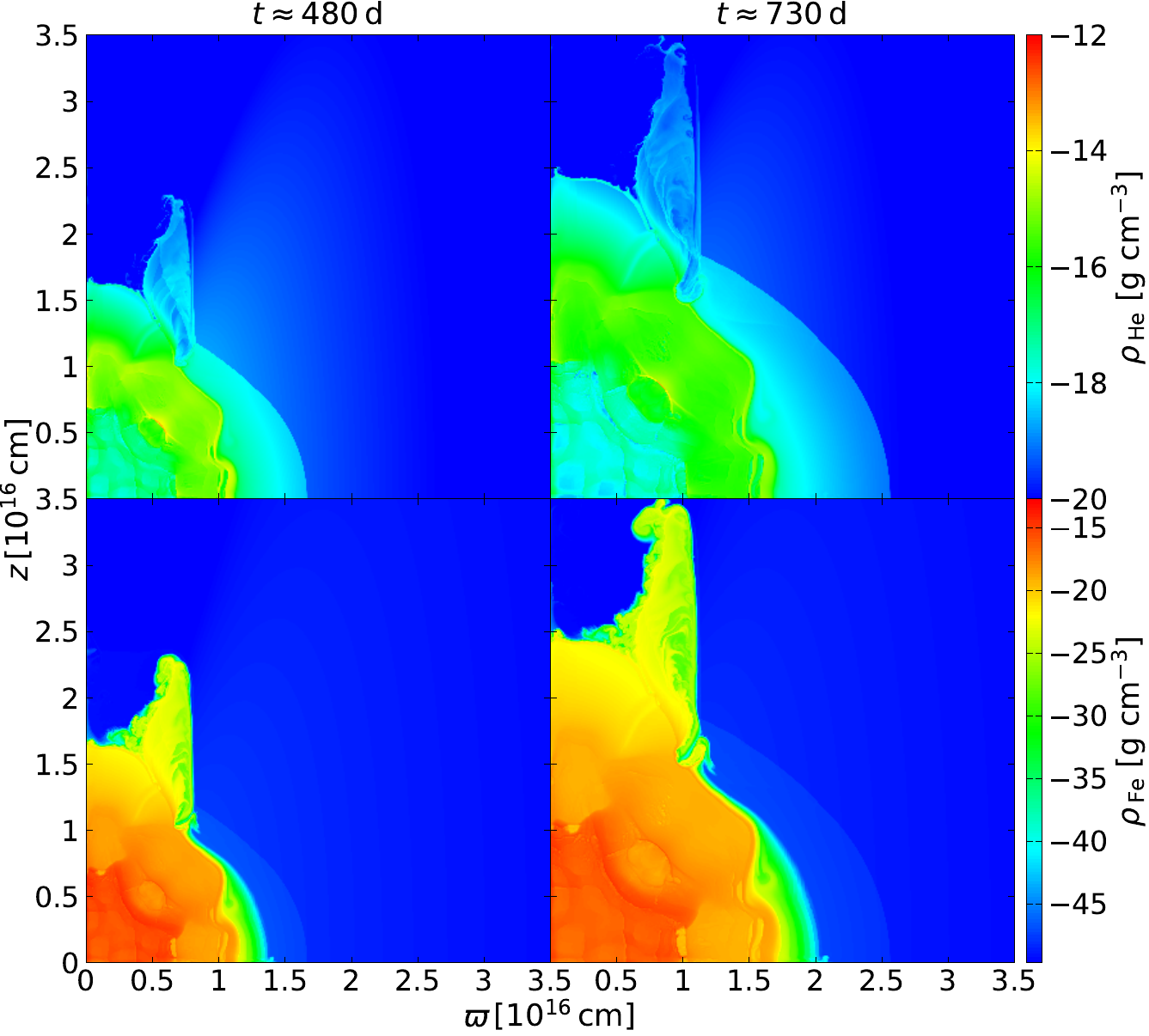}
    \hspace{0.8cm}\includegraphics[width=0.45\textwidth]{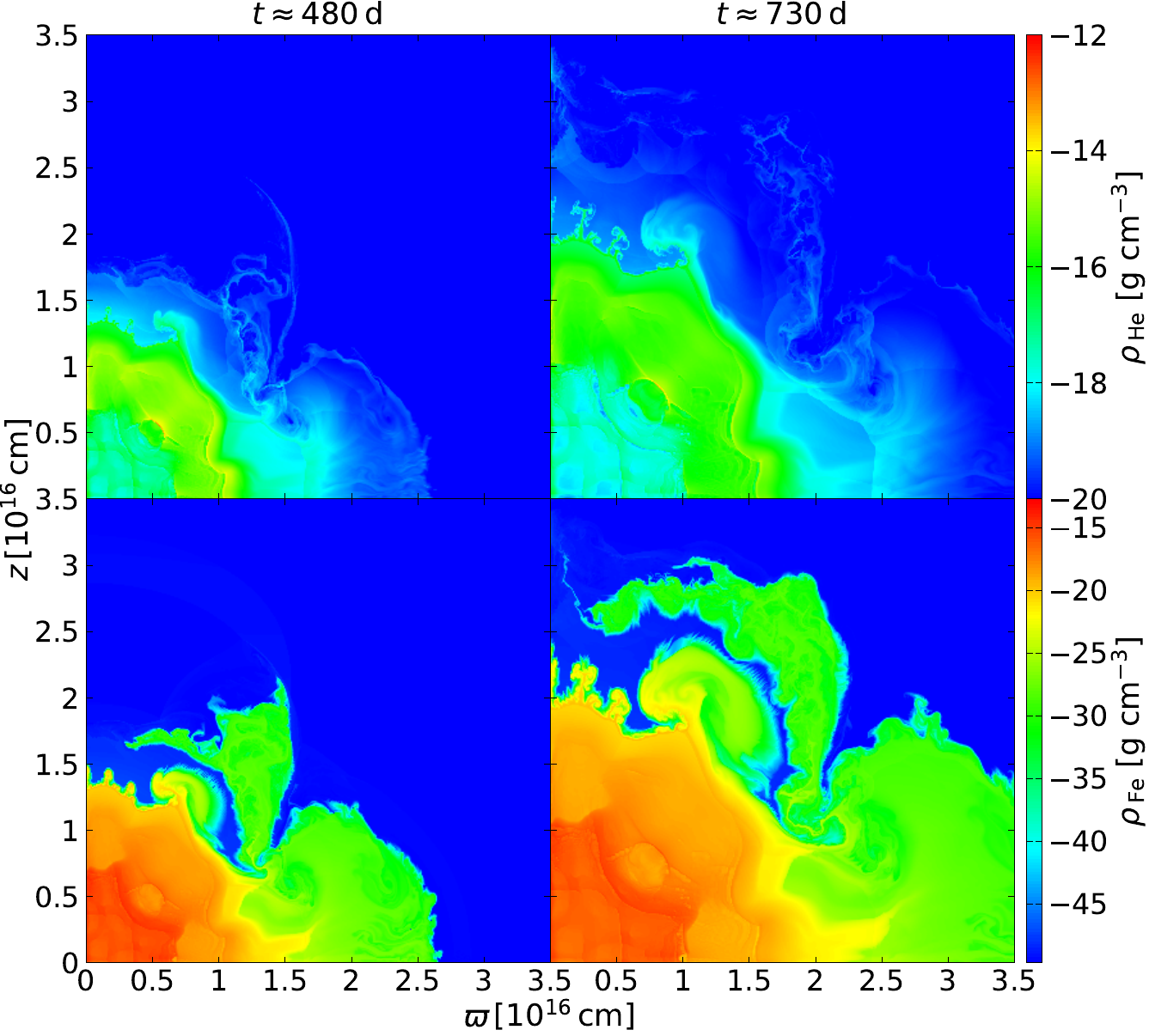
    }
    \caption{{\it\bf Left panel:} {\bf Model DMIN.} {\it Top row}: Evolution of He density at two evolved times marked in the graph. The density scaling is the same as in Fig.~\ref{disk_first}. {\it Bottom row}: Evolution of Fe density at the same two times with different scaling.
    {{\it\bf Right panel:} {\bf Model LMIN.} {\it Top row}: Example of map of the temporal evolution of helium density at two evolved times marked in the graph. The density scaling is the same as for the total density of this model in the top row of Fig.~\ref{lobes_first}. {\it Bottom row}: Map of the temporal evolution of iron density at the same two times, with significantly different density scaling.}
    \label{disk_lobes_helium_first}}
\end{figure*}
\section{Analytical scaling and parametrization of initial states of the models}
\label{analmodels}
\subsection{Initial scaling of spherically symmetric wind and circumstellar disc}
\label{appdisc}
We insert two components of the gaseous surroundings of an SN as an initial state: the spherically symmetric stellar wind, whose parameters correspond to a typical RSG stellar wind, and an aspherical component; that is, a circumstellar disc or a system of bipolar lobes, whose configuration may resemble the iconic Homunculus nebula around
the star $\eta$~Car \citep[e.g.,][]{2010HiA....15..373G,smith13_etacar}. Since we do not include the star's rotation, we do not refer to the disc as equatorial; other designations of directions, such as equatorial or polar, are therefore more or less formal. 
We set the stellar wind density, velocity, and temperature structure as 
\begin{align}
\rho_\text{w}=\rho_{0,\text{w}}\left(\frac{R_\star}{r}\right)^2,\quad v_\text{w}=15\,\text{km}\,\text{s}^{-1},\quad T_\text{w}=T_{\star,\text{eff}}\left(\frac{R_\star}{r}\right)^{0.75}.
\label{iniwind}
\end{align}
Here, $\rho_{0,\text{w}}\approx 10^{-15}\,\text{g}\,\text{cm}^{-3}$ is the wind base density (the density at the star's surface) that corresponds to the star-wind mass loss rate $\dot M_w=10^{-6}\,M_\odot\,\text{yr}^{-1}$ \citep[cf.][\citetalias{PK20}]{vanmarle10}, $r$ is the spherical radial distance from the star's centre, $v_\text{w}$ corresponds to the asymptotic RSGs wind velocity \citep{goldman17}, and the stellar (surface) effective temperature 
$T_{\star,\,\text{eff}}\approx 3300\,\text{K}$, 
which is typical for RSGs (all the star-subscripted quantities hereafter refer to the stellar surface values).

We include the initial gas-dynamic structure of the circumstellar $\alpha$-disk \citep{1973A&A....24..337S}, using the wind temperature profile in Eq.~\eqref{iniwind} for the disc in the following way \citep[cf.][]{1991MNRAS.250..432L,kurfurst2015thesis,2014A&A...569A..23K,2018A&A...613A..75K},
\begin{align}
\rho_\text{d}=\rho_{0,\text{d}}\left(\frac{R_\star}{\varpi}\right)^3\!\!\sqrt{\frac{r}{\varpi}}\,\,\text{exp}\left[-2\sqrt{\frac{R_\star}{\varpi}}\left(\frac{R_\star}{H_\star}\right)^2\left(1-\sqrt{\frac{\varpi}{r}}\right)\right],
\label{inidisk}
\end{align} 
where $\varpi$ denotes the distance from the disc axis ($r=\sqrt{\varpi^2+z^2}$. Here, $z$ is the vertical coordinate, $H=c_\text{s}/\Omega$ is the disc vertical scale height \citep{2011A&A...527A..84K}, $c_\text{s}(r)$ being the isothermal speed of sound related to the disc temperature $T_\text{d}(r)\propto r^{-0.75}$ (as in Eq.~\eqref{iniwind}) via the ideal gas EOS $c_\text{s}^2=kT/\mu m_u$ (where $k$ is the Boltzmann constant, $\mu$ is the mean molecular weight, and $m_u$ is the atomic mass unit), $\Omega(\varpi)=\sqrt{GM_\star/\varpi^3}$ being the
(Keplerian) angular velocity of the disc elements, and $P_\text{d}(r)=\rho_\text{d}\,c_\text{s}^2(r)$ is the 
isothermal ideal gas disc pressure. Since we assume a dominant external source of heating from the central star, 
we neglect the disc's internal viscous-heating \citep{2018A&A...613A..75K}. 
In the initial state setup, we neglect both velocity components, radial and azimuthal; the latter only formally defines $\Omega(\varpi)$ in the density prescription in Eq.~\eqref{inidisk}. We insert two different initial values of the disc mid-plane base mass density $\rho_{0,\text{d}}$ (i.e., the density at the disc mid-plane near the star's surface), $5\times 10^{-13}\,\text{g}\,\text{cm}^{-3}$ in model DMIN (about 10 times higher than in \citetalias{PK20}), and $5\times 10^{-11}\,\text{g}\,\text{cm}^{-3}$ in model DMAX. The $\rho_{0,\text{d}}$ value for model DMIN corresponds to the disc accretion (or decretion) rate $\dot M\approx 10^{-10}\,M_\odot\,\text{yr}^{-1}$, while model DMAX represents the disc accretion rate two orders of magnitude higher $\dot M\approx 10^{-8}\,M_\odot\,\text{yr}^{-1}$. The total mass of the spherical wind is in both cases $M_\text{w}\approx 10^{-3}\,M_\odot$, while the superposed initial mass of the disc and the underlying wind is $M_\text{w+d}\approx 4\times 10^{-1}\,M_\odot$ in the DMIN model and $M_\text{w+d}\approx 2.4\,M_\odot$ in the DMAX model, which is about $2$\textendash$3$ orders of magnitude higher
than that of the spherical wind alone. The parameters of both disc models were chosen to compare two adequately strong SN-CSM interactions with significantly different densities of a similar non-spherical formation.

We set the initial structure of the underlying stellar wind similar to models DMs while
 using the polytropic approximation and the ideal gas law for the temperatures of the expanding lobes in models LMs. We set the temperature in the unperturbed CSM to $20\,\text{K}$.
\subsection{Initial scaling of bipolar lobes}
\label{applobes}
For clarity, we list here Eqs.~(1)\,\textendash\,(7) (in a rather simplified form used in this work) and table 1 of \citet{2010MNRAS.402.1141G} used for the setup of bipolar lobe models, referred to in Sect.~\ref{rhdinicalc}. The initial profile of bipolar lobes in models LMs cannot be described simply analytically; we implement the semi-analytically pre-calculated structure that corresponds to five fundamental components (five "colliding winds" with different density and velocity parameters) of the Homunculus nebula, starting to evolve at five subsequent phases: pre-outburst wind (pew), giant eruption (ge), first post-outburst wind (pow3), minor eruption (me), and second post-outburst wind (pow5). We also paraphrase the models used in \citetalias{PK20} when we reduce the size of the nebula in a self-similar way to fit our current computational grid domain. For each of the five components $j$, a co-latitudinal velocity $v_j$ dependence is scaled as 
\begin{align}\label{gonzal1}
v_j=v_{1,j}F_j(\theta)\quad\text{with}\quad F_j (\theta)=\frac{\left(v_{2,j}/v_{1,j}\right)+\text{e}^{2z}}{1+\text{e}^{2z}},
\end{align}
where $\theta$ is the polar angle ($\theta=0$ at the pole, $\theta=\pi/2$ at the equator), $z_j=\lambda_j\cos\left(2\theta\right)$, and the parameters $v_{1,j}$, $v_{2,j}$ (velocities related to the polar
and equatorial directions, respectively) and $\lambda_j$ are given in Tab.~\ref{tablobes}. The initial mass density of the components is
\begin{align}
\rho_j=\rho_0\left(\frac{r_0}{r}\right)^2\frac{1}{F_j(\theta)}, 
\end{align}
where $\rho_0=\dot{m}_0/(4\pi v_0 r_0^2)$ is the density at the reference radius $r_0$ (selected as $\approx 8\times 10^{14}\,\text{cm}\approx 10\,R_\star$ here), where the downstream velocity $v_0=250\,\text{km}\,\text{s}^{-1}$ at $r_0$ is adopted as a free parameter, and 
$\dot{m}_0=10^{-5}\,M_\odot\,\text{yr}^{-1}$ is the total mass loss rate of the wind in the model LMIN. Besides that, we parametrize the mass loss rates of eruptive events in the model LMIN as $\dot{m}_\text{ge}=5\times 10^{-3}\,M_\odot\,\text{yr}^{-1}$ and $\dot{m}_\text{me}=10^{-4}\,M_\odot\,\text{yr}^{-1}$. The parameters $\dot{m}_0$, $\dot{m}_\text{ge}$, and $\dot{m}_\text{me}$ are in the models LMOD and LMAX increased by $10^2$, respectively (see Tab.~\ref{tablobes}). When the great eruption begins (at phase 2), the wind parameters increase to $av_0$ and $b\dot m_0$, with $a$ and $b$ being constants. The flow parameters form a working surface at the base of the wind. It corresponds to the large Homunculus lobe (formed in the "ge" phase) that expands with a constant
velocity 
\begin{align}
v_\text{ws}=\sigma v_0,
\end{align}
with $\sigma=(a^{1/2}+ab^{1/2})/(a^{1/2}+b^{1/2})$. The wind velocity is intermediate between the low-velocity downstream wind $v_0$ and the faster
upstream outflow $av_0$.
We have also set five time limits: $t_1=25\,\text{yr}$, $t_2=50\,\text{yr}$, $t_3=87.5\,\text{yr}$, $t_4=100\,\text{yr}$, and $t_5=112.5\,\text{yr}$, to separate five different courses of the nebula expansion behaviour. The two time intervals $\Delta t_2=30\,\text{yr}$ and $\Delta t_4=10\,\text{yr}$ represent the duration of the post-eruption
phases. The mass loss rates $\dot{m}_3$ and $\dot{m}_5$ in the phases after the giant and minor eruptions, respectively, are 
\begin{align}
\dot{m}_3 = \dot{m}_0\,\phi(t)
\end{align}
and
\begin{align}
\dot{m}_5 = \dot{m}_0\,\varphi(t),
\end{align}
where we adopt two functions
\begin{align}
\phi(t)=\frac{\dot{m}_\text{ge}}{\dot{m}_0}+\left(1-\frac{\dot{m}_\text{ge}}{\dot{m}_0}\right)\sin\left[\frac{\pi}{2}\left(\frac{t-t_2}{\Delta t_2}\right)\right]
\end{align}
and
\begin{align}
\varphi(t)=\frac{\dot{m}_\text{me}}{\dot{m}_0}+\left(1-\frac{\dot{m}_\text{me}}{\dot{m}_0}\right)\sin\left[\frac{\pi}{2}\left(\frac{t-t_4}{\Delta t_4}\right)\right],
\end{align}
which, due to the fact that different $\dot m$ form fractions, remain the same for all LMs models (see Tab.~\ref{tablobes}).
The computational domain is initially filled with a homogeneous
medium with a temperature of $T_a = 10^2\,\text{K}$ and a density of $n_a = 10^{-25}\,\text{g}\,\text{cm}^{-3}$. This is followed by the injection of a pre-outburst wind with a mass-loss rate of $\dot m_0$ at a distance of $r_0$,
with a temperature of $T_0=10^4\,\text{K}$. Afterwards, a gaseous distribution is gradually formed within the time limits according to the prescriptions and parameters introduced in this section.

\begin{table*}
\centering
\begin{threeparttable}
\caption{Parameters of the bipolar lobe outflows ($v_\text{p}$ and $v_\text{e}$ are the polar ($\theta=0$) and equatorial ($\theta=\pi/2$) velocities):\tnote{1}}
\label{tablobes}
\setlength{\tabcolsep}{7.25pt}
\def\arraystretch{1.12}
\begin{tabular}{lcccccc}
\hline
Phase ($j$)& $\lambda$  & $v_1\,\left(\text{km}\,\text{s}^{-1}\right)$ & $v_2\,\left(\text{km}\,\text{s}^{-1}\right)$
   & $v_\text{p}\,\left(\text{km}\,\text{s}^{-1}\right)$  & $v_\text{e}\,\left(\text{km}\,\text{s}^{-1}\right)$ & $\dot m\,\left(M_\odot\,\text{yr}^{-1}\right)$\\\hline Pre-outburst wind
 & $2.4$ & $250.00$ & $14.00$ & $248.07$ & $15.93$ & $10^{-5}$\\
Great eruption
 & $1.9$ & $687.76$ & $102.65$ & $674.95$ & $115.45$ & $5\times 10^{-3}$\\
 Post-outburst wind
 & $1.9$ & $500.00$ & $14.00$ & $ 489.37$ & $24.63$ & $10^{-5}\times\phi(t)$\\
Minor eruption & $1.9$ & $200.00$ & $10.00$ & $195.84$ & $14.16$ & $10^{-4}$\\
Post-outburst wind & $1.9$ & $500.00$ & $300.00$ & $495.62$ & $304.38$ & $10^{-5}\times\varphi(t)$\\
\hline
\end{tabular}
\label{tabgonzo}
\begin{tablenotes}
\item[1] {\footnotesize Values $\dot m$ for  are specified for the model LMIN. Values of constants in $\dot m$ for the models LMOD and LMAX increase always by a factor $10^2$, respectively, while the functions $\phi$ and $\varphi$ remain unchanged.}
\end{tablenotes}
\end{threeparttable}
\end{table*} 

\section{Animation documentation}
\label{animasect}\vspace{0.5cm}
(1) Animated version of the model DMIN, corresponding to Fig.~\ref{disk_first}, is linked here as \href{https://www.youtube.com/watch?v=iPgcKXu0LcI}{movie DMIN}.\vspace{0.5cm}\\
(2) Animated version of the model DMAX, corresponding to Fig.~\ref{disk_second}, is linked here as \href{https://youtu.be/xa84RQQwg8M}{movie DMAX}.\vspace{0.5cm}\\
(3) Animated version of the model LMIN, corresponding to Fig.~\ref{lobes_first}, is linked here as \href{https://youtu.be/FDEFx-JAEdw}{movie LMIN}.\vspace{0.5cm}\\
(4) Animated version of the model LMOD, corresponding to Fig.~\ref{lobes_second}, is linked here as \href{https://youtu.be/yQDR17Q6qZo}{movie LMOD}.\vspace{0.5cm}\\
(5) Animated version of the model LMAX, corresponding to Fig.~\ref{lobes_third}, is linked here as \href{https://youtu.be/lkK3WmZuYks}{movie LMAX}.

\end{appendix}

\end{document}